 \documentclass[final,5p,times,twocolumn,authornum]{elsarticle}

\usepackage{lipsum}
\usepackage{amsmath}
\usepackage{amsthm}
\usepackage{bm}% bold math
\usepackage{longtable}
\usepackage{natbib}
\usepackage{textcase}
\usepackage{url}
\usepackage{relsize}
\usepackage{amsfonts}
\usepackage{amsmath}
\usepackage{amssymb,epsf}
\usepackage{latexsym}
\usepackage{graphicx,epsfig}
\usepackage{float}
\usepackage{subfigure}
\usepackage{epstopdf}
\usepackage[colorlinks=true,citecolor=blue,linkcolor=blue,urlcolor=black]{hyperref}
\usepackage{psfrag}
\usepackage{wrapfig}
\usepackage{makeidx}
\usepackage{epsf}
\usepackage{xcolor}
\usepackage{mathtools}

\newtheorem{proposition}{Proposition}
\journal{Physics Letters B}

\begin{document}

\begin{frontmatter}



\title{Scalar--tensor baryogenesis: a scalar–tensor completion of gravitational baryogenesis}

\author[first,second]{David S. Pereira}\ead{djpereira@fc.ul.pt}
\affiliation[first]{organization={Departamento de Física, Faculdade de Ciências da Universidade de Lisboa},
            addressline={Edifício C8}, 
            city={Campo Grande},
            postcode={1749-016 Lisboa}, 
            %state={},
            country={Portugal}}
\affiliation[second]{organization={Instituto de Astrofísica e Ciências do Espaço},%Department and Organization
            addressline={Edifício C8}, 
            city={Campo Grande},
            postcode={1749-016 Lisboa}, 
            %state={},
            country={Portugal}}
\begin{abstract}
We propose \emph{Scalar--Tensor Baryogenesis} (STB), in which the $C\!P$-violating bias needed for baryogenesis is sourced by the \emph{gravitational} scalars that appear in scalar--tensor representations of modified gravity. Derivative couplings
$M_\ast^{-d}\nabla_\mu f(\phi_i)\,J^\mu_{B-L}$ act as an effective chemical potential $\mu_{B-L}\propto\dot f$ in an FRW background, driving the plasma to a nonzero equilibrium $B\!-\!L$ density while $B\!-\!L$-violating reactions are active. The asymmetry freezes in at the dynamically determined decoupling temperature $T_D$ fixed by $\Gamma_{B-L}(T_D)=H(T_D)$, giving $n_b/s\propto[\dot f/(M_\ast^d T)]_{T_D}$ up to sphaleron conversion. A key structural result is an explicit \emph{on-shell/background} map---through the Legendre relations defining the scalar potential---between curvature-based geometric Gravitational baryogenesis operators and their scalar--tensor counterparts, together with a canonical Einstein-frame description closely paralleling spontaneous/quintessential baryogenesis, but with a gravitational (not ad hoc matter) biasing field. The map is not a mere change of variables: it imposes consistency conditions (existence of the scalar--tensor branch, local invertibility of the Legendre map, and validity of the spectator regime), thereby restricting the admissible operator space and tying $\mu_{B-L}\propto\dot f$ to the modified-gravity dynamics once $F$ is specified. As an illustration, we implement STB in $F(R)=R^{1+\varepsilon}$ with $B\!-\!L$ violation from the dimension-five Weinberg operator, and reproduce the observed baryon asymmetry for $\varepsilon=\mathcal{O}(10^{-6})$ with $T_D\simeq 8.5\times10^{13}\,\mathrm{GeV}$ and negligible backreaction, while satisfying nucleosynthesis bounds and keeping the expansion arbitrarily close to the GR radiation solution.

\end{abstract}




\end{frontmatter}

%\tableofcontents



%%%%%%%%%%%%%%%%%%%%%%%%%%%%%%%%%%%%%%%%%%%%%%%%%%
\section{Introduction}\label{sec:intro}
%%%%%%%%%%%%%%%%%%%%%%%%%%%%%%%%%%%%%%%%%%%%%%%%%%
The observed asymmetry between matter and antimatter remains one of the most significant and persistent challenges in theoretical physics~\cite{c75ffd80-cea5-30a0-aee9-19091a4f0a9f,Barrow:2022gsu}. This asymmetry is quantified by the baryon-to-photon ratio $ \eta $, as determined by cosmological observations~\cite{Burles:2000ju,Burles:2000zk,WMAP:2003ivt}:  
\begin{equation}  
\eta \equiv \frac{n_B - n_{\bar{B}}}{n_\gamma} =  
\begin{cases}  
[5.8 - 6.6] \times 10^{-10}, & \text{(BBN)} \\  
(6.09 \pm 0.06) \times 10^{-10}, & \text{(CMB)}  
\end{cases}  
\end{equation}  
where $ n_B $, $ n_{\bar{B}} $, and $ n_\gamma $ denote the number densities of baryons, antibaryons, and photons, respectively. Equivalently, the baryon asymmetry is often expressed as the ratio $ n_b/s $, where $ n_b = n_B - n_{\bar{B}} $ is the net baryon number density and $ s $ is the entropy density:  
\begin{equation}  
\frac{n_b}{s} \equiv \frac{n_B - n_{\bar{B}}}{s} = \frac{\eta}{7.04} \, .  
\end{equation}  
For this work, we adopt the value $ \frac{n_b}{s} = (8.8 \pm 0.6) \times 10^{-11} $. Combined efforts from particle physics and gravitational physics have provided a deeper understanding of the mechanisms responsible for the asymmetry creation entitled baryogenesis mechanisms. Among the various proposed mechanisms, gravitational baryogenesis (GB)~\cite{Davoudiasl:2004gf} is particularly notable for its utilization of the gravitational interaction to generate the matter-antimatter asymmetry. The central feature of this mechanism is the interaction term~\cite{Davoudiasl:2004gf}
\begin{equation}\label{GB asymmetry}
    \mathcal{L}= \frac{1}{M_\ast^2} \left( \partial_\mu R J^\mu\right) \,,
\end{equation}
where $M_\ast$ represents the cutoff energy scale and $J^\mu$ being any current that leads to net $B-L$ charge in equilibrium. This term is conjectured to arise from higher-dimensional operators in supergravity frameworks and quantum gravity theories if $M_\ast$ is of the order of the reduced Planck scale~\cite{Davoudiasl:2004gf}. The direct coupling between the derivative of the Ricci scalar and $J^\mu$ leads to a dynamical violation of CPT symmetry in an expanding universe (for studies and constraints on CPT violations in the context of baryogenesis we refer to~\cite{Carroll:2005dj,Li:2006ss,Li:2008tma,Xia:2008si,Mavromatos:2013vqa,Mavromatos:2013boa,McDonald:2014yfg,Mavromatos:2017gyn,Zhai:2020vob}) leading to an asymmetry proportional to the time derivative of the Ricci scalar. Numerous studies have explored gravitational baryogenesis within the context of modified gravity theories~\cite{Li:2004hh,Lambiase:2006dq,Lambiase:2012tn,MohseniSadjadi:2007qk,Bhattacharjee:2020jfk,Mojahed:2024yus,Jaybhaye:2023lgr,Baffou:2018hpe,Nozari:2018ift,Sahoo:2019pat,Pereira:2024kmj,Odintsov:2016hgc,Pereira:2025flo,Cruz:2025fuk,Oikonomou:2016jjh,Odintsov:2016apy}. These modifications effectively address challenges such as generating a non-zero baryon asymmetry during the radiation-dominated epoch~\cite{Davoudiasl:2004gf} and mitigating instabilities arising from implementing Eq.\eqref{GB asymmetry} in the gravitational sector~\cite{Arbuzova:2023rri}.

In this work, we propose a mechanism denominated \emph{Scalar-Tensor Baryogenesis}, a hybrid mechanism combining elements of Gravitational and Spontaneous Baryogenesis. This mechanism not only recovers the original interaction term of Gravitational baryogenesis but also extends and generalizes this mechanism in the context of modified theories of gravity. We explore Scalar-tensor Baryogenesis in the context of scalar-tensor $F(R)$ gravities. 

The paper is organized as follows. In Sec.~\ref{sec:Scalar-tensor gravity} we briefly review scalar--tensor gravity and summarize how generic modified-gravity actions written in a ``geometric'' form can be mapped into an equivalent scalar--tensor representation, emphasizing the role of the Legendre-transform relations that connect the scalar potential to the underlying geometric invariants. In Sec.~\ref{sec:GB with STT} we introduce Scalar--Tensor Baryogenesis, derive the corresponding baryon-asymmetry yield, and clarify how the standard gravitational-baryogenesis operator and its common generalizations are recovered as particular limits within our framework, including a discussion of the Einstein-frame formulation and its relation to quintessential-type constructions. In Sec.~\ref{sec:STB-applied} we illustrate the mechanism in a representative $F(R)$ scenario with a minimal $B\!-\!L$-violating sector to demonstrate that the observed asymmetry can be reproduced within the phenomenologically allowed parameter space. Finally, Sec.~\ref{sec:Summuary and discussion} summarizes our main results and outlines potential directions for extending STB to other modified-gravity models and early-Universe histories.

%%%%%%%%%%%%%%%%%%%%%%%%%%%%%%%%%%%%%%%%%%%%%%%%%%%%%%%%%%%%%%%%%%%%%
\section{Scalar-tensor gravity}\label{sec:Scalar-tensor gravity}
%%%%%%%%%%%%%%%%%%%%%%%%%%%%%%%%%%%%%%%%%%%%%%%%%%%%%%%%%%%%%%%%%%%%%
In scalar–tensor theories, the gravitational action is formulated to describe the interplay between a scalar field and the metric tensor, thereby emphasizing the dual role of these fields in governing gravitational dynamics. First introduced by Brans and Dicke in 1961~\cite{Brans:1961sx}, this framework has since been generalized and is commonly written as~\cite{Will:2018bme}
\begin{equation}\label{actionSTT}
S=\frac{1}{2}\int \mathrm{d}^4x \sqrt{-g}\left( \phi R -\frac{\omega(\phi)}{\phi}\partial^\mu\phi\partial_\mu \phi - V(\phi) \right)
+S_\text{M}\left(g_{\mu\nu}, \Psi\right)\,,
\end{equation}
where the scalar field $\phi$ effectively replaces the usual gravitational constant, $\kappa = 8\pi G = M_{\rm Pl}^{-2}$ with $M_{\rm Pl}\simeq2.4\times 10^{18}\,\text{GeV}$. The function $\omega(\phi)$ denotes the coupling parameter regulating the scalar–gravity interaction, $V(\phi)$ is the scalar potential, and $S_\text{M}$ represents the matter action with matter fields $\Psi$. Teyssandier and Tourrenc~\cite{Teyssandier:1983zz} showed that $F(R)$ gravity can be reformulated as a scalar–tensor theory, often termed the scalar–tensor representation of $F(R)$ gravity. Starting from the action in the metric formalism
\begin{equation}
S= \frac{1}{2}\int \mathrm{d}^4 x \sqrt{-g}\  F(R) + S_\text{M}\left(g_{\mu\nu}, \Psi\right) ,
\end{equation}
one introduces an auxiliary field $\varphi$ to obtain~\cite{Teyssandier:1983zz,Sotiriou:2006hs}
\begin{equation}\label{stt F(R) 1}
S = \frac{1}{2}\int \mathrm{d}^4x \sqrt{-g} \left[F(\varphi) + F'(\varphi)(R-\varphi)\right] + S_\text{M}\left(g_{\mu\nu}, \Psi\right)\,.
\end{equation}

Although this form differs from the original $F(R)$ action, the equivalence is restored by imposing $R=\varphi$. Varying Eq.~~\eqref{stt F(R) 1} with respect to $\varphi$ yields $(R - \varphi) F''(\varphi) = 0$, which requires $F''(\varphi)\neq 0$ for consistency—ensuring the action is $R$-regular~~\cite{Magnano:1993bd}. Under this condition, redefining the field as $\phi=F'(\varphi)$ and the potential as $V(\phi) = \varphi(\phi)\phi - F(\varphi(\phi))$ gives
\begin{equation}\label{stt F(R)}
S = \frac{1}{2}\int \mathrm{d}^4x \sqrt{-g} \left(\phi R - V(\phi)\right) + S_\text{M}\left(g_{\mu\nu}, \Psi\right)\,,
\end{equation}
which corresponds to the scalar–tensor action \eqref{actionSTT} with vanishing kinetic term. Equivalently, if the $F(R)$ action is $R$-regular, the potential can be identified with the Legendre transform of $F(R)$, with $\phi = F'(R)$ and $R = V'(\phi)$, thereby reproducing the scalar–tensor form~\cite{OHanlon:1972xqa}.

The transition from the geometric to the scalar-tensor representation can be performed using either the metric or Palatini formalism, yielding distinct outcomes~\cite{Sotiriou:2006hs,Barrow:1988xh}. In this work, we focus only on the metric formalism, while the Palatini approach will be studied in future research. Furthermore, it is common in the literature to move from the Jordan frame to the Einstein frame by applying a conformal transformation of the metric~\cite{Dabrowski:2008kx}. 
By applying the conformal transformation
\begin{equation}\label{eq:conformal transf}
    g_{\mu\nu}=\Omega^{-2} \tilde{g}_{\mu\nu}=\frac{M_{\rm Pl}^2}{2\phi} \tilde{g}_{\mu\nu}\,,
\end{equation}
to Eq.~\eqref{stt F(R)} one obtains the action
\begin{equation}
    S = \int \mathrm{d}^4x \sqrt{-\tilde{g}}\left(\frac{M_{\rm Pl}^2}{2}\tilde{R} -\frac{1}{2}\partial_\mu \tilde{\phi} \partial^\mu \tilde{\phi} -\tilde{V}(\tilde{\phi})
    + \tilde{\mathcal{L}}_m\right)\,,
\end{equation}
where the " $\tilde{}$ " represents quantities in the Einstein frame with $\tilde{\mathcal{L}}_m=\Omega^{-4}\mathcal{L}_m$, $\tilde{V}(\tilde{\phi})$ is given by
\begin{equation}
    \tilde{V}(\tilde{\phi}) = {V}(\phi({\tilde{\phi}}))\Omega^{-4}\,,
\end{equation}
and $\tilde{\phi}$ defined as 
\begin{equation}\label{eq:phi Eframe}
    \textrm{d}\tilde{\phi} = \sqrt{\frac{3}{2}}M_{\rm Pl} \frac{\mathrm{d}\phi}{\phi}\,,
\end{equation}
that leads to
\begin{equation}\label{eq:new scalar EF}
    \tilde{\phi} = \sqrt{\frac{3}{2}}M_{\rm Pl}{\ln\left(\frac{\phi}{\phi_0}\right)}\,,
\end{equation}
with $\phi_0$ being a constant with units of $\text{mass}^2$ that comes from the integration of Eq.\eqref{eq:phi Eframe}.  This method is especially useful in models with a single geometric scalar, such as $F(R)$ gravity. However, since our main goal is to study the role of the scalar-tensor form of modified gravity in baryogenesis, the main results will be explored in the Jordan frame and the Einstein frame will be briefly studied for completeness.
%%%%%%%%%%%%%%%%%%%%%%%%%%%%%%%%%%%%%%%%%%%%%%%%%%%%%%%%%%%%%%%%
\subsection{General scalar-tensor representation}
%%%%%%%%%%%%%%%%%%%%%%%%%%%%%%%%%%%%%%%%%%%%%%%%%%%%%%%%%%%%%%%%%
The previous passage from the geometric representation to the scalar-tensor representation, done for $F(R)$, only involved one geometrical quantity, but one can generalize for $N$ geometrical scalar degrees of
freedom. 

Consider a general modified gravity theory involving $N$ geometric scalar degrees of freedom, where the action is given by
\begin{equation}\label{geometrical}
S = \frac{1}{2} \int \mathrm{d}^4 x \sqrt{-g} \, F\left(x_1, \ldots, x_N\right) + S_\text{M}\left(g_{\mu\nu}, \Psi\right),
\end{equation}
with $F(x_1, \ldots, x_N)$ representing a function dependent on the $N$ geometric scalar degrees of freedom, $x_i$. Analogous to the procedure used for $F(R)$ gravity, we now introduce $N$ auxiliary fields $\chi_i$, where each field corresponds to one of the geometric scalar degrees of freedom. This allows the action in Eq.~\eqref{geometrical} to be rewritten as
\begin{align}\label{stt geral 1}
S &= \frac{1}{2} \int \mathrm{d}^4 x \sqrt{-g} \left[F\left(\chi_1, \ldots, \chi_N\right) \right. \nonumber \\ 
&+\left. \sum_{i=1}^N \frac{\partial F}{\partial \chi_i}\left(x_i -\chi_i\right)\right] + S_\text{M}\left(g_{\mu\nu}, \Psi\right).
\end{align}

We now identify the true scalar fields of the theory, those that act as mediators of the gravitational interaction, alongside the metric tensor. These scalar fields are defined as
\begin{equation}\label{scalars stt}
\phi_i \equiv \frac{\partial F}{\partial \chi_i}(\chi_1,\ldots,\chi_N)\,,
\end{equation}
allowing to express the action in terms of non-minimal couplings between the scalar fields and their respective geometric degrees of freedom
\begin{equation}\label{stt geral}
S=\frac{1}{2} \int \mathrm{d}^4 x \sqrt{-g} \left[\sum_{i=1}^N \phi_i x_i - V\left(\phi_1, \ldots, \phi_N\right)\right] +S_\text{M}\left(g_{\mu\nu}, \Psi\right),
\end{equation}
where $V(\phi_1, \ldots, \phi_N)$ is the potential associated with the scalar fields defined as
\begin{align}
\label{eq:LegendreV}
V(\phi_1,\ldots,\phi_N)&\equiv \sum_{i=1}^{N}\phi_i\,\chi_i(\phi_1,\ldots,\phi_N) \\ \nonumber
&- F\!\left(\chi_1(\phi_1,\ldots,\phi_N),\ldots,\chi_N(\phi_1,\ldots,\phi_N)\right)\,.
\end{align}

The geometric and scalar-tensor representations of a given theory, are fundamentally equivalent, as both formulations describe the same underlying physical phenomena. From a mathematical point of view, one can relate both descriptions by varying the action \eqref{stt geral 1}
with respect to each auxiliary field $\chi_k$, resulting in the field equations
\begin{equation}
\sum_{j=1}^N \frac{\partial^2 F}{\partial\chi_k\,\partial\chi_j}\,(x_j-\chi_j)=0,
\qquad k=1,\ldots,N.
\end{equation}
that can be rewritten in a matrix form $M\mathbf{\Delta} = 0$, with $M_{kj}\equiv \frac{\partial^2 F}{\partial\chi_k\,\partial\chi_j}$ and $\Delta_j\equiv x_j-\chi_j$, as
\begin{equation}
M\mathbf{\Delta} = 
    \begin{bmatrix}
\frac{\partial^2 F }{\partial \chi_k\partial \chi_j}
\end{bmatrix}
\begin{bmatrix}
x_j-\chi_j
\end{bmatrix}=0\,,
\end{equation}
where $\frac{\partial^2 F }{\partial \chi_k\chi_j} = \frac{\partial^2 F }{\partial \chi_j\chi_k}$. This system is known to possess a unique solution if the determinant of the matrix $ M $ is non-zero. Ensuring that $\text{det}(M) \neq 0$, guarantees that the only unique solutions will be $ x_i = \chi_i $. Substituting these solutions into the action \eqref{stt geral} confirms that the equation reduces to the form of action \eqref{geometrical}, thereby demonstrating the equivalence between the two formulations and affirming the well-defined nature of the scalar-tensor representation. Additionally, it is crucial to note that this description holds for a general action only when the action does not involve a direct coupling term between the matter Lagrangian and some function, as is the case in nonminimally coupled $F(R)$ theories~\cite{Bertolami:2007gv}. Here, in contrast to the previous description done for the general case, besides the usual potential $V(\phi)$ there exists an additional potential, $U(\phi)$, responsible for the non-minimal coupling of the theory to the matter sector.
%%%%%%%%%%%%%%%%%%%%%%%%%%%%%%%%%%%%%%%%%%%%%%%%%%%%%%%%%%%%%%%%%%%%%
\section{Scalar-tensor Baryogenesis}\label{sec:GB with STT}
%%%%%%%%%%%%%%%%%%%%%%%%%%%%%%%%%%%%%%%%%%%%%%%%%%%%%%%%%%%%%%%%%%%%%
The passage from the geometric framework to the scalar-tensor representation introduces new scalar fields, which, in turn, open the possibility of introducing new interaction terms of the form of gravitational baryogenesis~\cite{Davoudiasl:2004gf}, spontaneous baryogenesis~\cite{Cohen:1987vi,Cohen:1988kt} and quintessential baryogenesis~\cite{DeFelice:2002ir}. Hence, one can consider for each of these scalar fields an interaction of the form 
\begin{equation}\label{scalar int term}
 \mathcal{L}_{\text{int}} = \frac{\mathrm{c}_{\text{STB}}}{M_\ast^d} \left( \nabla_\mu (\phi_j) J^\mu\right) \,,
\end{equation}
$\mathrm{c}_{\text{STB}}$ is a dimensionless coupling constant, and $d$ denotes the positive mass dimension of the scalar field $\phi_j$, $M_\ast$ represents the effective field theory cutoff, $\Lambda_{\text{EFT}} = M_\ast$. The effective description is therefore assumed to hold in the regime $T\ll M_\ast$ relevant for the epoch of interest, so that higher-order operators suppressed by additional powers of $M_\ast$ are negligible. The subscript $j$ labels a scalar field compatible with a coherent EFT formulation. The current $J^\mu$ may correspond to a baryonic current or any other current carrying a net baryon-minus-lepton $(B - L)$ charge in thermal equilibrium. For this work we couple to the anomaly–free $B\!-\!L$ current,
\begin{equation}
  J^\mu_{B\!-\!L}=\sum_i q_i\,j_i^\mu,\qquad q_i\equiv(B\!-\!L)_i,
\end{equation}
so that, after integrating by parts,
\begin{equation}
  \mathcal{L}_{\rm int}
  = -\frac{\mathrm{c}_{\text{STB}}}{M_\ast^{\,d}}\,\phi_j\,\nabla_\mu J^\mu_{B\!-\!L}
    + \frac{\mathrm{c}_{\text{STB}}}{M_\ast^{\,d}}\nabla_\mu(\phi_j J^\mu),
\end{equation}
and only $\nabla_\mu J^\mu_{B\!-\!L}$ matters (the last term is a boundary term).
This choice is convenient for two reasons. (i) Anomalies: in SM gauge backgrounds $\nabla_\mu J^\mu_{B-L}=0$, while $\nabla_\mu J^\mu_{B,L}\propto W^a_{\mu\nu}\tilde W^{a\mu\nu}$ up to a hypercharge–scheme term removable by a Bardeen counterterm, with $W^a_{\mu\nu}$ the SU(2)$_L$ field strength and $\tilde W^{a\mu\nu}\equiv \tfrac12\,\varepsilon^{\mu\nu\rho\sigma} W^a_{\rho\sigma}
$ its dual~\cite{Adler:1969gk,Bell:1969ts,Bardeen:1969md,Fujikawa:1979ay,Fujikawa:1980eg,tHooft:1976rip,tHooft:1976snw,Bodeker:2020ghk}. Hence $\phi_j\nabla_\mu J^\mu$ induces an axion-like term $\sim \phi_j W\tilde W$, whose relevance depends on the chosen $B$-violating dynamics~\cite{Fujikawa:1980eg,Harvey:2005it}. (ii) Sphalerons: electroweak sphalerons violate $B{+}L$ but conserve $B{-}L$; any surviving baryon asymmetry therefore tracks $B{-}L$~\cite{Kuzmin:1985mm,Harvey:1990qw}.

The interaction term~\eqref{scalar int term} leverages the extra degree of freedom introduced by a generic function $ F(x_1, \ldots, x_N) $ to couple with the baryonic current, potentially arising from the compactified low-energy limits of higher-dimensional theories or effective quantum gravity like Gravitational baryogenesis~\cite{Dabrowski:2008kx}.

%%%%%%%%%%%%%%%%%%%%%%%%%%%%%%%%%%%%%%%%%%%%%%%%%%%%%%
\subsection{Baryon asymmetry generation}\label{sec:Baryon asymmetry generation}
%%%%%%%%%%%%%%%%%%%%%%%%%%%%%%%%%%%%%%%%%%%%%%%%%%%%%%
To analyze how the interaction~\eqref{scalar int term} biases the thermal plasma and generates an asymmetry, we begin by evaluating the \emph{matter} Hamiltonian density—i.e., we Legendre–transform only the matter fields present in primordial plasma while treating the scalar $\phi_j(t)$ as a classical, homogeneous (cosmological) background because $\phi_j$ is a gravitational degree of freedom rather than a plasma field. In this setting the interaction contributes as
\begin{equation}
\mathcal{H}_{\rm int}\;=\;-\mathcal{L}_{\rm int}\,,
\end{equation}
and by using the homogeneity imposed by the Cosmological principle, which gives $\partial_i\phi_j=0$, $\partial_\mu\phi_j\,J^\mu_{B\!-\!L}=\dot\phi_j\,J^0_{B\!-\!L}$ with $J^0_{B-L}=n_{B\!-\!L}$ being the $B\!-\!L$ number density, it can be simplified to
\begin{equation}
\mathcal{H}_{\rm int} = -\frac{\mathrm{c}_{\text{STB}}}{M_\ast^{d}}\dot\phi_jn_{B\!-\!L}
\equiv  -\sum_i \mu_in_i,
\end{equation}
hence each species $i$ carrying $q_i$ experiences $\mu_i=q_i\,\frac{\mathrm{c}_{\text{STB}}\dot{\phi}_j}{M_\ast^d}$ and its particle/antiparticle energies shift by $\mp\,\mu_i$ (so $\bar\mu_i=-\mu_i$) \cite{Davoudiasl:2004gf,Cohen:1987vi,Cohen:1988kt,DeFelice:2002ir,Sadjadi:2007dx,DeFelice:2004uv,Kamada:2012ht}. Crucially, the time–dependent background $\dot\phi_j\neq0$ provides an \emph{effective CPT–violating} bias.
By itself this operator cannot change a strictly conserved charge:
if $\partial_\mu J^\mu_{B-L}=0$ it reduces to a boundary term.
Hence a genuine $B\!-\!L$–violating interaction is indispensable; when such processes are in equilibrium,
the bias $\mu$ drives the plasma to a nonzero equilibrium $B\!-\!L$ density.

A viable realization of the STB mechanism proceeds as follows. Let $\Gamma_{B-L}(T)$ denote the rate of a chosen $B\!-\!L$–violating process.
For $T>T_{\rm D}$ with $\Gamma_{B-L}\gg H$, the process remains in thermal equilibrium and continuously
establishes the biased asymmetry set by $\mu$.
As the Universe cools to the decoupling temperature $T_{\rm D}$ defined by $\Gamma_{B-L}(T_{\rm D})\approx H(T_{\rm D})$ the mechanism freezes out.
For $T<T_{\rm D}$, with $\Gamma_{B-L}\ll H$, the comoving asymmetry is thereafter conserved
(up to possible late entropy injection). To compute this asymmetry, we will use
\begin{equation}\label{eq:nbl density}
    n_{B\!-\!L}=\sum_i q_i \Delta n_i\,,
\end{equation}
where $\Delta n_i$ is the net density number of fermions, $n_{i}-\bar{n}_i$, that in the equilibrium regime is given by~\cite{Kolb:1990vq,Pereira:2024vdk}
\begin{equation}\label{net baryon}
    \Delta n_i = \frac{g_i}{2\pi^2} \int \mathrm{d}p\ p^2 \left[ \left(e^{ {(E(p) - \mu_i)}/{T}} + 1\right)^{-1} -\left(e^{{(E(p) + \mu_i)}/{T}} + 1\right)^{-1} \right] \,,
\end{equation}
where $g_i$ denotes the number of internal degrees of freedom for a given fermion. Assuming a homogeneous and isotropic universe and considering the limits $T \gg m$ and $T \gg \mu$, the above expression simplifies to
\begin{equation}
    \Delta n = \frac{g_i T^3}{6\pi^2} \left[ \pi^2 \left(\frac{\mu_i}{T} \right) + \left(\frac{\mu_i }{T}\right)^3 \right] \approx \frac{g_iT^2\mu_i}{6}\,,
\end{equation}
that in combination with Eq.~\eqref{eq:nbl density} gives
\begin{equation}\label{eq:nbl density2}
     n_{B\!-\!L} = \frac{\hat{g}\mathrm{c}_{\text{STB}}}{6 M_\ast^d} T^2 \dot{\phi}_j\,.
\end{equation}
where $\hat{g}\equiv \sum_i g_i q_i^2$ \footnote{If one wishes to include susceptibility effects~\cite{Bodeker:2020ghk,Comelli:1994di}, impose the fast-equilibrium and gauge-neutrality constraints (Yukawas, sphalerons, hypercharge neutrality), solve the $\mu_i$ in terms of $\mu_{B-L}$, and replace $\hat g=\sum_i g_i q_i^2$ by an effective $\hat g_{\rm eff}$ obtained from that solution. In what follows we adopt the simplest free-gas estimate for the charge susceptibility (i.e.\ $\hat g=\sum_i g_i q_i^2$), noting that imposing the full set of equilibrium and gauge-neutrality constraints at $T\simeq T_D$ would only rescale $\hat g$ by an $\mathcal{O}(1)$ factor and hence shift the inferred parameter values at the same level.} and in the SM with three families, no $\nu_R$, $\hat g=13$ (or $16$ if three right-handed neutrinos are included). Using now the entropy density $s=\frac{2\pi^2}{45}g_{\ast s} T^3$, the final asymmetry generated by the interaction term \eqref{scalar int term} will be given by 
\begin{equation}\label{asymmetry phi i}
    \frac{n_{B-\!\!L}}{s} \simeq \frac{15\hat{g}\mathrm{c}_{\text{STB}}}{4\pi^2 g_{\ast s}} \frac{\dot{\phi}_j}{M^d_\ast T} \Bigg|_{T_\text{D}} \,,
\end{equation}
where $g_\ast s$ is the total degrees of freedom for relativistic particles contributing to the entropy of the universe~\cite{Kolb:1990vq}. In thermal equilibrium we can consider $g_\ast s \simeq g_\ast$ where $g_\ast$ is the number of degrees of freedom for relativistic species with $g_\ast=106.75$ for relativistic particles ($T\gg 100 \ \text{GeV}$) in the SM. Finally, because we observe a baryon asymmetry, the required primordial $B\!-\!L$ is fixed by sphaleron equilibrium to \cite{Bodeker:2020ghk,Harvey:1990qw}
\begin{equation}
    \frac{n_b}{s}= c_s \frac{n_{B\!-\!L}}{s}\,,
\end{equation}
where $c_s= \frac{8N_f+4N_H}{22N_f+13N_H}$ with $N_{f}$ being the number fermion families and $N_{H}$ the Higgs doublets, resulting in $c_s=\frac{28}{79}$ for the SM.

One can also consider more complex interactions terms that are functions or combinations of the scalar fields of the theory leading to
\begin{equation}\label{general scalar int term}
\mathcal{L}_{\text{int}} = \frac{\mathrm{c}_{\text{STB}}}{M_\ast^d} \left( \nabla_\mu f(\phi_1,\ldots,\phi_N) J^\mu\right) \,,
\end{equation}
where $f(\phi_1,\ldots,\phi_N)$ represents a general function of the scalar fields or a single scalar field, possessing a mass dimension of $d$. This interaction term gives the asymmetry
\begin{equation}\label{asymmetry f phi i}
    \frac{n_b}{s} \simeq \frac{15\hat{g} \mathrm{c}_{\text{STB}} c_s}{4\pi^2 g_{\ast}} \frac{\partial_t(f(\phi_1,\ldots,\phi_N))}{M^d_\ast T} \Bigg|_{T_\text{D}} \,.
\end{equation}

%%%%%%%%%%%%%%%%%%%%%%%%%%%%%%%%%%%%%%%%%
\subsection{Einstein frame formalism}
%%%%%%%%%%%%%%%%%%%%%%%%%%%%%%%%%%%%%%%%%
The Einstein frame presents an interesting framework as in this frame one typically has a kinetic term in the action which brings new dynamics. As previously mentioned, the transition from the Jordan frame to the Einstein frame is primarily relevant in the context of $F(R)$ gravity. Therefore, the following analysis is carried out using the considerations $\phi_j = \phi$ and $f(\phi_1, \ldots, \phi_N) = f(\phi)$, with $\phi \equiv \frac{dF(R)}{dR}$.

Under the conformal transformation \eqref{eq:conformal transf}, a current $J^\mu$ generally transforms as\footnote{For instance, consider the current $J^\mu = B \bar{\psi} \gamma^\mu \psi$, where $B$ denotes the baryon number of a fermionic field $\psi$, and $\gamma^\mu$ are the gamma matrices in curved spacetime. These transform as $\tilde{\gamma}^\mu = \Omega^{-1} \gamma^\mu$, while the fermion field transforms as $\tilde{\psi} = \Omega^{-3/2} \psi$, and $\tilde{\bar{\psi}} = \Omega^{-3/2} \bar{\psi}$~\cite{Birrell:1982ix,Dabrowski:2008kx}, yielding $\tilde{J}^\mu = \Omega^{-4} J^\mu$.}
\begin{equation}
J^\mu = \Omega^4 \tilde{J}^\mu,
\end{equation}
while the volume element transforms as $\sqrt{-g} = \Omega^{-4} \sqrt{-\tilde{g}}$. The interaction term \eqref{general scalar int term}, which generalizes \eqref{scalar int term} depending on the functional form of $f$, transforms in the Einstein frame as
\begin{equation}
\int \mathrm{d}^4x \, \sqrt{-\tilde{g}} \, \frac{\mathrm{c}_{\text{STB}}}{M_\ast^2} \, \partial_\mu f\left( \phi_0 e^{\sqrt{\frac{2}{3}} \frac{\tilde{\phi}}{M_{\text{Pl}}}} \right) \tilde{J}^\mu.
\end{equation}

For example, choosing $f(\phi) = \alpha \sqrt{\frac{3}{2}} M_{\text{Pl}} \ln\left( \frac{\phi}{\phi_0} \right)$ with $\alpha$ a parameter of mass dimension one, the interaction reduces to
\begin{equation}\label{eq:STB EF}
\int \mathrm{d}^4x \, \sqrt{-\tilde{g}} \, \frac{\mathrm{c}_{\text{STB}}}{\tilde{M}_\ast} \, \partial_\mu \tilde{\phi} \, \tilde{J}^\mu,
\end{equation}
where $\tilde{M}_\ast \equiv M_\ast^2 / \alpha$. A natural and convenient choice is $\alpha = M_\ast$, ensuring that $M_\ast$ remains the EFT cutoff scale, a convention adopted henceforth. Notably, in the Einstein frame, STB closely parallels quintessential baryogenesis~\cite{DeFelice:2002ir}: for the chosen functional form of $f$, the interaction term is identical. Equation~\eqref{eq:STB EF} makes the connection to spontaneous/quintessential baryogenesis manifest: for suitable choices of the functional coupling $f(\phi)$, the Einstein-frame interaction reduces to the familiar derivative coupling $\partial_\mu \tilde{\phi}\,\tilde{J}^\mu$.  This is precisely why STB is nontrivial: although the Einstein-frame operator matches the canonical spontaneous/quintessential form, the potential $\tilde V(\tilde\phi)$ and trajectory $\tilde\phi(t)$ are not model-building inputs but are fixed by the Jordan-frame modified-gravity Lagrangian through the Legendre structure. In this sense STB turns derivative baryogenesis into a controlled probe of the gravity sector rather than an adjustable scalar-sector mechanism. Concretely, in standard spontaneous/quintessential baryogenesis one typically chooses a scalar potential and background evolution $\tilde{\phi}(t)$ largely independently of the gravitational sector. In STB, by contrast, $\tilde{\phi}$ (equivalently $\phi_i$ in the Jordan frame) is tied to the underlying geometric invariants through the Legendre-transform relations, e.g.\ $x_i=V_{\phi_i}$ (and $R=V_\phi$ in $F(R)$ models). In addition, STB in the Einstein-frame formulation is well-defined and canonical, meshes smoothly with quintessential-type constructions, making it easier to embed various mechanisms such as inflation, reheating, and related dynamics typically developed in the EF.

Moreover, the transformation to the Einstein frame must be analyzed with care. Although the Einstein-frame rewriting can bring new phenomenology, the conformal transformation induces non-minimal couplings between the matter sector and the scalar field. Hence, the $B\!-\!L$ (or $B$) interaction, the associated chemical-potential interpretation, and other intermediate quantities are modified in form and are not to be compared term-by-term with their Jordan-frame counterparts as separately conformal-invariant objects. Therefore, the relevant statement is not term-by-term invariance of intermediate quantities, but the equivalence of the final physical prediction when the complete transformed system (gravity, matter sector, and freeze-out condition) is treated consistently. In this sense, the Einstein-frame expressions presented should be understood as the Einstein-frame representation of the same underlying mechanism, rather than as an independent frame-dependent prescription. In the present work, the final baryon asymmetry is interpreted in the Jordan frame (where matter is minimally coupled and baryons follow free-fall geodesics).

%%%%%%%%%%%%%%%%%%%%%%%%%%%%%%%%%%%%%%%%%%%%%%%%%
\subsection{Relation between gravitational baryogenesis and  scalar-tensor baryogenesis}\label{sec:RelationGBSTB}
%%%%%%%%%%%%%%%%%%%%%%%%%%%%%%%%%%%%%%%%%%%%%%
Due to the very definition of the gravitational scalars in the scalar--tensor representation, it is natural that STB and generalized GB are related. In the ST formulation the scalars are introduced as
$\phi_i\equiv \partial F/\partial\chi_i$ in Eq.~\eqref{scalars stt}; on the auxiliary-field solution one has $\chi_i=x_i$, so that locally $\phi_i=\phi_i(x_1,\ldots,x_N).$ Thus, treating $\phi_i$ as a composite function of the geometric invariants gives
\begin{equation}
\label{eq:chainrule_phiJ}
\nabla_\mu\phi_i\,J^\mu_{B-L}
=\sum_{j=1}^N F_{ij}(x)\,\nabla_\mu x_j\,J^\mu_{B-L}\,,
\end{equation}
where $F_i\equiv \partial F/\partial x_i$ and $F_{ij}\equiv \partial^2 F/(\partial x_i\,\partial x_j)$ is the Hessian of the geometric
Lagrangian (evaluated on the branch $\chi_i=x_i$). More generally, for the STB interaction \eqref{general scalar int term} (with $f$ differentiable), one has the explicit ``pull-back'' to geometric derivatives,
\begin{equation}
\label{eq:chainrule_fphi_2}
\mathcal{L}_{\text{int}}
=\frac{c_{\rm EFT}}{M_\ast^{\,d}}\sum_{i=1}^N\sum_{j=1}^N {f}_{\phi_i}(\phi_1,\ldots,\phi_N)\,F_{ij}(x_1,\ldots,x_N)\,\nabla_\mu x_j\,J^\mu_{B-L}\,.
\end{equation}

Equations \eqref{eq:chainrule_phiJ}--\eqref{eq:chainrule_fphi_2} make explicit that, whenever the scalar--tensor branch exists and the map $\phi_i(x)$ is well defined, STB operators correspond \emph{locally} to linear combinations of generalized GB operators $\nabla_\mu x_j\,J^\mu_{B-L}$, with coefficients constrained by the underlying modified-gravity function $F$ (through its Hessian) and by the chosen STB coupling ${f}$ (through ${f}_{\phi_i}$).

At the same time, Eqs.~\eqref{eq:chainrule_phiJ}--\eqref{eq:chainrule_fphi_2} should not be read as an off-shell equivalence between STB and GB, nor as evidence that STB is merely ``GB with a more complicated functional form.'' Firstly, the chain-rule rewriting is only available after restricting to the scalar--tensor branch (auxiliary-field solution $\chi_i=x_i$ and local invertibility of the map $\phi_i(x)$), so it is a local/background statement rather than an identity at the level of the action. Secondly, even when the rewriting exists, the resulting geometric combination is \emph{not freely specifiable}: the relative weights multiplying $\nabla_\mu x_j\,J^\mu_{B-L}$ are fixed and correlated by the modified-gravity sector through the Hessian $F_{ij}$ and by the chosen STB coupling through $f_{\phi_i}$, instead of being arbitrary EFT functions. With these distinctions in mind, we now make precise (i) how the operator classes commonly employed in generalized GB admit an on-shell/background scalar--tensor representation, and (ii) under which local conditions scalar--tensor couplings can be pulled back to geometric operators.

\subsubsection{GB $\to$ STB map}
In modified-gravity implementations of gravitational baryogenesis (often termed
\emph{generalized baryogenesis}), one extends the original operator
$\nabla_\mu R\,J^\mu$ to derivative couplings built from geometric scalar invariants
$x_i$ entering the gravitational Lagrangian $F(x_1,\ldots,x_N)$. A representative class is
\begin{equation}
\label{eq:GB_geo_class}
\mathcal{L}_{\rm int}^{\rm (geo)}=\frac{c_{\rm EFT}}{M_\ast^{\,d}}\,
\nabla_\mu \mathcal{I}(x_1,\ldots,x_N)\,J^\mu_{B-L}\,,
\end{equation}
where $c_{\rm EFT}$ is a coupling constant like $c_{\text{STB}}$ and $\mathcal{I}$ is differentiable (e.g.\ $C^1$) and has mass dimension $d$.

Whenever the theory admits the scalar--tensor representation reviewed in
Sec.~\ref{sec:Scalar-tensor gravity} and matter is minimally coupled, variation of the
\emph{gravity sector} with respect to $\phi_i$ yields the Legendre relations
\begin{equation}
\label{eq:Legendre_relations}
x_i = V_{\phi_i}(\phi_1,\ldots,\phi_N)\qquad (i=1,\ldots,N)\,,
\end{equation}
which may be used \emph{along a given homogeneous background solution} to translate
\eqref{eq:GB_geo_class} into scalar--tensor variables. This is the sense in which generalized
GB operators admit a scalar--tensor representation: it is a background/on-shell rewriting, not an
off-shell identity at the level of the action.

\begin{proposition}[On-shell/background dictionary for GB$\rightarrow$STB operators] \label{cl:dictionary} Let a modified-gravity theory with action \eqref{geometrical} admit a scalar--tensor representation \eqref{stt geral} such that $x_i=\chi_i$ follows from the auxiliary-field equations (equivalently, $\det M\neq 0$ in the region of interest), and assume that the Legendre map is locally invertible so that $V(\phi_1,\ldots,\phi_N)$ is well defined. Then any `geometric'' baryogenesis interaction built from the invariants $x_i$ of the form \begin{equation}\label{eq:geo_dictionary_general} \mathcal{L}_{\rm int}^{\rm (geo)}=\frac{\mathrm{c}_{\text{EFT}}}{M_\ast^{\,d}}\, \nabla_\mu \,\mathcal{I}(x_1,\ldots,x_N)\,J^\mu_{B-L}\,, \end{equation} admits the on-shell/background rewriting \begin{equation} \mathcal{L}_{\rm int}^{\rm (geo)} \;\mathrel{\overset{\text{on-shell}}{=}}\; \frac{\mathrm{c}_{\text{EFT}}}{M_\ast^{\,d}}\, \nabla_\mu \,\mathcal{I}\!\bigl(V_{\phi_1},\ldots,V_{\phi_N}\bigr)\,J^\mu_{B-L}\,, \label{eq:geo_dictionary_onshell} \end{equation} i.e.\ it can be expressed in scalar--tensor language with $f(\phi_1,\ldots,\phi_N)\equiv \mathcal{I}(V_{\phi_1},\ldots,V_{\phi_N})$ which constitutes the interaction term~\eqref{general scalar int term}. Conversely, a scalar--tensor coupling of the form \eqref{general scalar int term} can be pulled back to a geometric operator whenever the inverse map $\phi_i=\phi_i(x_1,\ldots,x_N)$ exists in the regime of interest, by defining $\mathcal{I}(x)\equiv f(\phi(x))$. In particular, for $\mathcal{I}(x)=x_i$ one recovers \begin{equation} \nabla_\mu x_i\,J^\mu_{B-L} \;\;\mathrel{\overset{\text{on-shell}}{=}}\;\; \nabla_\mu\bigl(V_{\phi_i}\bigr)\,J^\mu_{B-L}\,, \label{eq:geo_to_st_basic} \end{equation} so that standard gravitational baryogenesis $\nabla_\mu R\,J^\mu$ is obtained as the special case $x_i=R$ in $F(R)$ models with $R=V_\phi$. 
\end{proposition}

\subsubsection{STB $\to$ GB map}
The map STB $\rightarrow$ GB is more subtle. Because in STB one writes derivative couplings directly in terms of the gravitational scalars, including genuinely multi-field functional combinations as presented in Eq.~\eqref{general scalar int term}, unlike \eqref{eq:GB_geo_class}, these interactions depend explicitly on $\phi_i$ and therefore source the scalar equations of motion. To make this explicit in the $N$-field setting, consider the representative STB coupling to a single scalar field $\phi_j$, \begin{equation}\label{general scalar int term_j} \mathcal{L}_{\text{int}} = \frac{\mathrm{c}_{\text{STB}}}{M_\ast^d}\, \nabla_\mu f(\phi_j)\, J^\mu_{B-L}\,, \end{equation} with $f$ differentiable and of mass dimension $d$ (the case $f(\phi_j)=\phi_j$ reproduces Eq.~\eqref{scalar int term}). Integrating by parts (discarding a boundary term) gives \begin{equation}\label{eq:int_by_parts_here} \sqrt{-g}\,\mathcal{L}_{\rm int} \;\doteq\; -\sqrt{-g}\,\frac{\mathrm{c}_{\text{STB}}}{M_\ast^{\,d}}\,f(\phi_j)\,\nabla_\mu J^\mu_{B-L}\,, \end{equation} and varying the \emph{total} scalar--tensor action (gravity sector \eqref{stt geral} plus the STB interaction above) with respect to $\phi_k$ then yields the sourced Legendre-type equations \begin{equation}\label{eq:sourced_legendre_here} 
x_k - V_{\phi_k} = \frac{2\,\mathrm{c}_{\text{STB}}}{M_\ast^{\,d}}\, \delta_{kj}\,f_{\phi_j}(\phi_j)\;\nabla_\mu J^\mu_{B-L}\,, \qquad (k=1,\ldots,N)\,, \end{equation} 
so that only the equation for the directly coupled scalar ($k=j$) is sourced.\footnote{If $f$ depends on several fields, $\delta_{kj}f_{\phi_j}$ is replaced by $f_{\phi_k}$.} This previous result appears to break the capability of recasting the STB interaction terms in terms of the gravitational baryogenesis interaction ones~\eqref{eq:GB_geo_class}. However, for baryogenesis, it is standard and sufficient to treat $\mathcal{L}_{\rm int}$ as a small EFT perturbation that biases the thermal plasma while inducing negligible backreaction on the homogeneous cosmological background. Concretely, we compute the background evolution from the minimally coupled modified-gravity sector \eqref{stt geral} and evaluate the chemical potential and the yield at leading order in $\mathrm{c}_{\text{STB}}/M_\ast^{\,d}$ on that background. In this spectator approximation, the gravity-sector Legendre relations may be consistently imposed along the background trajectory, 
\begin{equation} x_i(t)=V_{\phi_i}(\phi(t))\qquad\text{(background, leading order in $\mathrm{c}_{\text{STB}}/M_\ast^{\,d}$)}\,, \end{equation} 
so that scalar-tensor operators~\eqref{general scalar int term} can be rewritten \emph{on-shell} as $\mathcal{I}(x_1,\ldots,x_N)$ for the purpose of computing the baryon yield. A sufficient condition for the validity of this approximation is that the interaction-induced source terms in \eqref{eq:sourced_legendre_here} remain small compared to the leading gravity-sector terms during the epoch of interest, e.g. \begin{equation}\label{eq:spectator_condition_here} \left| \frac{2\,\mathrm{c}_{\text{STB}}}{M_\ast^{\,d}}\, f_{\phi_k}\,\nabla_\mu J^\mu_{B-L} \right| \ll \left|V_{\phi_k}\right| \qquad(\text{for each relevant }k)\,. 
\end{equation}

In this approximation one may consistently impose \eqref{eq:Legendre_relations} along the background
trajectory to translate between geometric and scalar--tensor operator choices when computing $\dot f$ and
the baryon yield.

\begin{proposition}[STB$\rightarrow$GB pull-back]
\label{prop:STBtoGB}
Assume that in the regime of interest the inverse map $\phi_i=\phi_i(x_1,\ldots,x_N)$ exists locally. Then the STB operator \eqref{general scalar int term} admits the local pull-back
\begin{equation}
\label{eq:STB_to_GB_pullback}
\mathcal{L}_{\rm int}^{\rm (STB)}
\;\mathrel{\overset{\rm bg/on\text{-}shell}{=}}\;
\frac{c_{\rm STB}}{M_\ast^{\,d}}\,
\nabla_\mu \mathcal{I}(x_1,\ldots,x_N)\,J^\mu_{B-L}\,,
\end{equation}
with the equality understood locally in field space (and, in our applications, evaluated along the spectator
background).
\end{proposition}

Propositions~\ref{cl:dictionary} and \ref{prop:STBtoGB} show that STB is not a mere re-labelling of generalized GB, but a
\emph{model-tied} framework that organizes derivative baryogenesis operators. In generalized GB one typically postulates a geometric
functional $\mathcal{I}(x_i)$ and studies the induced bias; in STB the bias is built from the gravitational scalars of the
scalar--tensor completion and is therefore correlated with the modified-gravity sector through the Legendre structure. In particular,
once a modified-gravity Lagrangian $F(x_1,\ldots,x_N)$ is specified, the scalar potential---and hence the dynamics controlling
$\mu_{B-L}\propto \dot f$---is fixed, turning baryogenesis into a probe of the gravity model rather than an arbitrary EFT choice.
Moreover, the GB$\to$STB map makes explicit when a geometric baryogenesis operator admits a consistent scalar--tensor origin,
yielding \emph{selection rules} on generalized-GB couplings and clarifying when the correspondence holds only on a particular branch
and/or locally in field space. The corresponding limitation---made explicit by the propositions---is that the geometric/scalar--tensor
dictionary is intrinsically \emph{local} and \emph{branch-dependent}, requiring the existence of the scalar--tensor representation and
the relevant invertibility conditions; in STB this is a feature, since it renders the consistency requirements of a proposed operator
choice explicit.

Beyond this structural organization, STB highlights extensions that are not naturally captured by purely geometric parametrizations.
First, geometric invariants and gravitational scalars need not be paired one-to-one: through the Legendre relations a given invariant may bias baryogenesis through a \emph{different} scalar degree of freedom, and genuinely multi-field couplings $f(\phi_1,\ldots,\phi_N)$ allow cross-coupled directions in field space whose geometric pull-back can be non-manifest (or only local/branch-dependent). This opens the door to richer bias dynamics---e.g.\ cancellations or sign changes in $\mu_{B-L}\propto\dot f$, bounded choices of $f$ that regulate the bias at high temperature, or multi-step freeze-out when distinct $B\!-\!L$ channels decouple at different epochs.
Second, because STB couplings depend explicitly on gravitational scalars, they generically source the scalar equations of motion and can deform the naive gravity-sector Legendre relations; backreaction is therefore an \emph{internal} and controllable effect, enabling systematic corrections to the yield and to $T_D$ within the same EFT expansion.
Third, STB provides a sharp organizing principle for EFT consistency: the existence of the scalar--tensor branch, local invertibility of the Legendre map, and the spectator bound act as \emph{selection rules} identifying which geometric operator choices admit a consistent scalar--tensor completion in the regime of interest.
Fourth, STB admits a canonical Einstein-frame description in which the interaction can reduce to the standard derivative coupling $\partial_\mu\tilde\phi\,\tilde J^\mu$; this offers a convenient language to embed STB into realistic early-Universe histories (inflation, reheating and possible intermediate epochs), where the relevant temperature controlling the bias is often $T_{\max}$ rather than $T_{\rm RH}$ and where a modified $T(t)$ relation shifts the decoupling condition $\Gamma_{B-L}=H$, with calculable dilution from late entropy injection.
Finally, since the same scalar--tensor sector controls the effective Planck mass and/or the expansion history, STB naturally invites correlated probes beyond $n_b/s$: modifications to the propagation and transfer of primordial or stochastic gravitational waves provide a particularly interesting target, and (if the biasing scalar fluctuates) isocurvature constraints may also become relevant. A quantitative exploration of these inflation/reheating and gravitational-wave connections is left for future work.

%%%%%s%%%%%%%%%%%%%s%%%%%%%%%%%%%s%%%%%%%%%%
\section{Application of STB}\label{sec:STB-applied}
%%%%%%%%%%%%%%%%%%%%%%%%%%%%%%%%s%%%%%%%%%%%%%s%%%%%%%%%%%%%s%%%%%%%%%%
In this section we illustrate the scalar--tensor baryogenesis framework of
Sec.~\ref{sec:GB with STT} in a simple and widely used $F(R)$ benchmark. We will use its scalar field $\phi\equiv\frac{dF(R)}{dR}$ to build the interaction term
\begin{equation}\label{eq:phi asymmetry}
\mathcal{L}_{\phi} \;=\; \frac{\mathrm{c}_{\text{STB}}}{M_\ast^2}\,\partial_\mu \phi\, J^\mu\,,
\end{equation}
that leads to the asymmetry
\begin{equation}\label{eq:asymmetry phi}
\frac{n_b}{s} \;\simeq\; \,\frac{15\,\hat g\,\mathrm{c}_{\text{STB}}\,c_s}{4\pi^2 g_\ast}\,
\left.\frac{\dot\phi}{M_\ast^2\,T}\right|_{T_D}\,,
\end{equation}

To make the discussion concrete we adopt the power-law $F(R)$ model
\begin{equation} \label{eq:modelF(R)}
F(R) = M_{\text{Pl}}^{2-2\varepsilon} R^{1+\varepsilon}\,, 
\end{equation} 
with $\varepsilon>0$ that reduces to the Einstein--Hilbert form in the limit $\varepsilon\to 0$ and has the potential 
\begin{equation}\label{eq:potentialF(R)} V(\phi) = \varepsilon (1+\varepsilon)^{-\frac{1+\varepsilon}{\varepsilon}} M_{\text{PL}}^{2-\frac{2}{\varepsilon}} \phi^{\frac{1+\varepsilon}{\varepsilon}}\,. 
\end{equation}

This model has become a standard benchmark in $F(R)$ phenomenology and has been explored across a broad range of contexts, including cosmological dynamics and exact solutions \cite{Carloni:2005Rn,CapozzielloDeFelice:2008Noether}, cosmological perturbations and growth constraints \cite{ParkHwangNoh:2011Scaling}, early-Universe bounds from big-bang nucleosynthesis \cite{Kusakabe:2015yaa}, Solar-System tests \cite{Zakharov:2006SSRn}, galactic rotation-curve phenomenology \cite{CapozzielloCardoneTroisi:2007LSB,FrigerioMartinsSalucci:2007RC}, and compact-star applications \cite{Capozziello:2015yza}. For general overviews of $F(R)$ gravity and its observational probes, see also \cite{DeFelice:2010aj,SotiriouFaraoni:2010RMP}.

\subsection{Field Equations and Cosmological Dynamics for Baryogenesis}
Varying the scalar--tensor action with respect to the metric yields the field equations
\cite{Goncalves:2021vci}
\begin{equation}\label{eq:fields}
\phi R_{\mu\nu}
-\frac{1}{2}g_{\mu\nu}\bigl(\phi R - V(\phi)\bigr)
-\bigl(\nabla_\mu\nabla_\nu-g_{\mu\nu}\Box\bigr)\phi
= T_{\mu\nu}\,,
\end{equation}
where $\Box\equiv\nabla^\alpha\nabla_\alpha$. Variation with respect to $\phi$ gives the
Legendre-transform relation
\begin{equation}\label{eq:Vphi}
R = V_\phi\,,
\end{equation}
and, since the matter sector is minimally coupled in $F(R)$ gravity, the Bianchi identities imply the
standard conservation law
\begin{equation}
\nabla_\mu T^{\mu\nu}=0\,.
\end{equation}

We consider a spatially flat FLRW metric,
\begin{equation}\label{eq:metric}
{\rm d}s^{2}=-{\rm d}t^{2}+a^{2}(t)\,{\rm d}V^2,
\end{equation}
and a perfect-fluid matter sector,
\begin{equation}\label{EM tensor}
T_{\mu\nu}=(\rho+p)u_\mu u_\nu +p g_{\mu\nu}\,,
\end{equation}
with barotropic equation of state $p=w\rho$. Under the assumption of homogeneity, the cosmological
field equations reduce to the modified Friedmann and acceleration equations
\begin{equation}\label{00}
\dot{\phi}H + \phi H^2  = \frac{1}{3}\rho + \frac{1}{6} V(\phi)\,,
\end{equation}
\begin{equation}\label{ii}
\phi(2\dot{H}+3H^2)+2H\dot{\phi}=-w\rho+ \frac{1}{2}V(\phi) - \ddot{\phi}\,,
\end{equation}
together with
\begin{equation}\label{VphiFLRW}
V_{\phi}=R=6(\dot{H} + 2H^2)\,,
\end{equation}
and the standard continuity equation
\begin{equation}\label{conservation stt F(R) cosmo}
\dot{\rho}+3H\rho(1+w)=0.
\end{equation}

As in GR, these equations are not all independent: Eq.~\eqref{ii} follows from Eq.~\eqref{00} together with \eqref{VphiFLRW} and \eqref{conservation stt F(R) cosmo} by virtue of the Bianchi identities. Accordingly, one may work with the reduced set \eqref{00}, \eqref{VphiFLRW} and
\eqref{conservation stt F(R) cosmo}, supplemented by $p=w\rho$. To close the system we fix: (i) the equation-of-state parameter $w=1/3$ consistent with the radiation dominated era, (ii) the gravitational model via the potential $V(\phi)$, and (iii) the background expansion history through the power-law ansatz
\begin{equation}\label{scale factor}
a(t)\propto t^{\eta},
\end{equation}
with constant $\eta$, so that $H=\eta/t$ and $\dot H=-\eta/t^2$. This consideration allows to compute $\phi$ by using Eq.~\eqref{eq:potentialF(R)} and Eq.~\eqref{VphiFLRW} yielding
\begin{equation}\label{eq:phiF(R)}
\phi(t)=\phi_0\,t^{-2\varepsilon}\,,
\end{equation}
and therefore
\begin{equation}\label{scalar_F(R)}
\dot{\phi}(t)= -2\varepsilon\,\phi_0\,t^{-2\varepsilon-1}\,,
\end{equation}
where
\begin{equation}
\phi_0=\bigl[6(2\eta^2-\eta)\bigr]^{\varepsilon}(1+\varepsilon)\,M_{\rm Pl}^{\,2-2\varepsilon}\,.
\end{equation}

This explicitly exhibits the key point for the present mechanism: even for $\varepsilon\ll 1$ the geometric scalar has $\dot\phi\neq 0$, thereby sourcing the effective chemical potential in Eq.~\eqref{eq:phi asymmetry} while the background expansion remains arbitrarily close to the GR
radiation solution.

Substituting these expressions into the cosmological field equation \eqref{00}, one obtains
\begin{equation}
    \rho(t)=3M_{\rm Pl}^{2-2\varepsilon}\Bigl(6(2\eta^2-\eta)\Bigr)^{\varepsilon} \left[(1-\varepsilon)\eta^2-\varepsilon(1+2\varepsilon)\eta\right]t^{-2\varepsilon-2}\,.
\end{equation}

On the other hand, the conservation law \eqref{conservation stt F(R) cosmo} implies a power-law behavior for the energy density,
\begin{equation}\label{rho_F(R)}
    \rho(t)=\rho_0\,t^{-4\eta}\,.
\end{equation}
so consistency between the time dependence in the two expressions for \(\rho(t)\) then fixes
\begin{equation}
    \eta=\frac{1+\varepsilon}{2}\,,
\end{equation}
and determines the normalization as
\begin{equation}
    \rho_0 = 3M_{\rm Pl}^{2-2\varepsilon}\Bigl(6(2\eta^2-\eta)\Bigr)^{\varepsilon} \left[(1-\varepsilon)\eta^2-\varepsilon(1+2\varepsilon)\eta\right]\,.
\end{equation}

One can relate the time with the temperature by using the equation that relates the total radiation density with the energy of all relativistic species
\begin{equation}\label{eq:rho_stat}
    \rho = \frac{\pi^2 g_\ast}{30} T^4\,,
\end{equation}
and the form for $\rho(t)$ obtained from the cosmological equations, Eq.~\eqref{rho_F(R)}, giving
\begin{equation}\label{eq:tvsT}
    t = \left(\frac{30\rho_0}{\pi^2g_\ast T^4} \right)^{\frac{1}{2\varepsilon+2}}\,.
\end{equation}

\subsection*{$B\!-\!L$ violation and the decoupling temperature}
The decoupling temperature $T_D$ should not be regarded as a free parameter ({in contrast with what is often done in the gravitational-baryogenesis literature, where $T_D$ is frequently treated as an external input}). It is fixed by the competition between the microscopic $B\!-\!L$-violating reaction rate and the Hubble expansion rate being determined by the freeze-out condition $\Gamma_{B\!-\!L}(T_D)=H(T_D)$ which relates $T_D$ uniquely to the specified $B\!-\!L$ sector and the expansion history.

A minimal and well-motivated realization of the required $B\!-\!L$ violation, that we will adopt, is the dimension–five Weinberg operator in the SM EFT \cite{Weinberg:1979sa,Turner:2018mwh,Pascoli:2018cqk},
\begin{equation}\label{eq:Weinberg}
\mathcal{L}_\text{W}=-\frac{\lambda_{\alpha\beta}}{\Lambda} \ell_{\alpha L}^{i} \varepsilon^{ij} H^j C \ell_{\beta L} ^k \varepsilon^{kl} H^l +\text{h.c},
\end{equation}
where $\ell_L = (\nu_L, l_L)^T$ in the $SU(2)_L$ gauge space, $\lambda_{\alpha\beta}=\lambda_{\beta\alpha}$ are effective Yukawa couplings with flavour indices $\alpha,\beta=e,\mu,\tau$ and $C$ is the charge conjugation matrix. In the early Universe it mediates $\Delta L=2$ scatterings such as $LL\leftrightarrow HH$ (and crossed channels), yielding a thermally averaged interaction rate that can be expressed directly in terms of the light-neutrino masses as \cite{Weinberg:1979sa,Turner:2018mwh,Pascoli:2018cqk}
\begin{equation}
\label{eq:GammaW}
\Gamma_W(T) =\frac{3}{4\pi^{3}}\,
\frac{ \bar{m}_{\nu}^{2}}{v^{4}_H}\;T^{3}\,,
\end{equation}
where $\bar m_\nu^2\equiv \sum_{i=1}^3 m_{\nu_i}^2$ is the sum of the light-neutrino mass eigenvalues squared, and after electroweak symmetry breaking the Weinberg operator yields the Majorana mass matrix $(m_\nu)_{\alpha\beta}=\lambda_{\alpha\beta}v_H^2/\Lambda$ and $v_H=246$ GeV is the Higgs Vacuum Expectation Value (VEV).

Employing the condition $\Gamma_W(T_D) = H(T_D)$ in combination with the last equation, Eq~\eqref{eq:tvsT} and $H=\eta/t$ gives the decoupling temperature 
\begin{equation}
    T_D = \left[\frac{2\pi^3 v^4_H (1+\varepsilon)}{3\bar{m}_\nu^2} \left(\frac{\pi^2 g_\ast}{30\rho_0}\right)^{\frac{1}{2\varepsilon+2}}\right]^{\frac{1+\varepsilon}{1+3\varepsilon}}\,.
\end{equation}

 Furthermore, with the previous condition established it is easy to prove that for $T\gg T_D$ $\Gamma_W\gg H$ and for $T<T_D$ $\Gamma_W<H$, fundamental conditions to guarantee the viability of the STB mechanism. Imposing the decoupling condition $\Gamma_W(T_D)=H(T_D)$, the equilibrium status of the $\Delta L=2$ processes is controlled by the ratio
\begin{equation}
\frac{\Gamma_W(T)}{H(T)}
=\frac{\Gamma_W(T_D)}{H(T_D)}
\left(\frac{T}{T_D}\right)^{\frac{1+3\varepsilon}{1+\varepsilon}}
=\left(\frac{T}{T_D}\right)^{\frac{1+3\varepsilon}{1+\varepsilon}}\,,
\end{equation}
hence, because $\varepsilon>-1$, the exponent is positive and therefore $\Gamma_W/H\gg1$ for $T\gg T_D$, while $\Gamma_W/H<1$ for $T<T_D$. This guarantees that $B\!-\!L$ violation is efficient above $T_D$ and shuts off after decoupling, as required for the STB freeze-out picture.

\subsubsection*{Consistency of the spectator approximation}
As stressed in Sec.~\ref{sec:RelationGBSTB}, the dictionary $x_i=V_{\phi_i}$ is to be used
\emph{on-shell/background}, i.e.\ in a spectator regime where $\mathcal{L}_{\rm int}$ biases the plasma but does not
backreact on the homogeneous solution. Since the STB coupling depends explicitly on the gravitational scalar, it also
sources the $\phi$ equation of motion. For the benchmark interaction \eqref{eq:phi asymmetry}, variation of the full action gives
\begin{equation}\label{eq:phi_sourced_check}
R - V_\phi \;=\;\frac{2\,\mathrm{c}_{\text{STB}}}{M_\ast^{2}}\,\nabla_\mu J^\mu_{B-L}\,.
\end{equation}

The spectator approximation (background $R=V_\phi$) is therefore consistent provided the RHS is small compared to the
curvature scale. We quantify this by
\begin{equation}\label{eq:delta_STB_def}
\delta_{\rm STB}(T)\equiv
\frac{\left|\frac{2\,\mathrm{c}_{\text{STB}}}{M_\ast^{2}}\,\nabla_\mu J^\mu_{B-L}\right|}{|R|}\,,
\qquad \delta_{\rm STB}\ll 1\,.
\end{equation}

In a homogeneous FRW background, $\nabla_\mu J^\mu_{B-L}=\dot n_{B-L}+3Hn_{B-L}$. With $B\!-\!L$ violation mediated by the
Weinberg operator, the near-equilibrium evolution is well captured by
\begin{equation}\label{eq:boltzmann_relax}
\dot n_{B-L}+3H n_{B-L}\;=\;-\Gamma_W(T)\,\Bigl[n_{B-L}-n_{B-L}^{\rm eq}\Bigr]\,.
\end{equation}

In the near-equilibrium regime relevant for baryogenesis ($\Gamma_W\gg H$), the solution of
\eqref{eq:boltzmann_relax} is rapidly attracted to $n_{B-L}^{\rm eq}$ after a short transient. From
\eqref{eq:boltzmann_relax} one has identically
\begin{equation}\label{eq:div_bound_general}
\bigl|\nabla_\mu J^\mu_{B-L}\bigr|
=\Gamma_W(T)\,\bigl|n_{B-L}-n_{B-L}^{\rm eq}\bigr|\,.
\end{equation}

To obtain a conservative upper bound on the interaction-induced backreaction, we may take the maximal
departure compatible with this relaxation form, $\bigl|n_{B-L}-n_{B-L}^{\rm eq}\bigr|\;\lesssim\;\bigl|n_{B-L}^{\rm eq}\bigr|
\Rightarrow
\bigl|\nabla_\mu J^\mu_{B-L}\bigr|\;\lesssim\;\Gamma_W(T)\,\bigl|n_{B-L}^{\rm eq}\bigr|$ that by using the equilibrium expression \eqref{eq:nbl density2} for $n_{B-L}^{\rm eq}$ then yields, for general $T$,
\begin{equation}\label{eq:delta_general_T}
\delta_{\rm STB}(T)\;\lesssim\;
\frac{\hat g\,\mathrm{c}_{\text{STB}}^{\,2}}{3}\,
\frac{\Gamma_W(T)\,T^{2}\,|\dot\phi(T)|}{M_\ast^{4}\,|R(T)|}\,.
\end{equation}

For the background solution $a\propto t^\eta$ with
$\eta=(1+\varepsilon)/2$, Eq.~\eqref{VphiFLRW} gives $R=6(\dot H+2H^2)=\frac{12\varepsilon}{1+\varepsilon}\,H^2$ and from Eq.~\eqref{eq:phiF(R)} one finds $ |\dot\phi|=\frac{2\varepsilon}{t}\phi=\frac{4\varepsilon}{1+\varepsilon}\,\phi H$. Substituting these previous results into \eqref{eq:delta_general_T} yields the compact
general-$T$ estimate
\begin{equation}\label{eq:delta_general_compact}
\delta_{\rm STB}(T)\;\lesssim\;
\frac{\hat g\,\mathrm{c}_{\text{STB}}^{\,2}}{9}\,
\left(\frac{\phi(T)}{M_\ast^{2}}\right)\left(\frac{T}{M_\ast}\right)^2
\left(\frac{T}{T_D}\right)^{\frac{1+3\varepsilon}{1+\varepsilon}}\,.
\end{equation}

Equation~\eqref{eq:delta_general_compact} must only hold over the temperature range where the EFT description is intended to apply,
namely $T\in[T_D,T_{\max}]$, where $T_{\max}$ is the maximal temperature attained after reheating (or, more generally, the maximal
temperature for which the EFT with cutoff $M_\ast$ is trusted). Since $\varepsilon>-1$ the exponent in
$\Gamma_W/H=(T/T_D)^{\frac{1+3\varepsilon}{1+\varepsilon}}$ is positive, and the bound in \eqref{eq:delta_general_compact} grows
monotonically with $T$ (up to slow $\mathcal{O}(1)$ variations in $\phi$). Therefore it is sufficient to impose the spectator condition
at the upper end of the interval:
\begin{equation}
\delta_{\rm STB}(T)\le \delta_{\rm STB}(T_{\max})\ll 1
\qquad\text{for all}\qquad T\in[T_D,T_{\max}]\,,
\end{equation}

The EFT itself requires $T_{\max}\ll M_\ast$, so the factor $(T_{\max}/M_\ast)^2$ provides strong suppression. Moreover, writing
\begin{equation}\label{eq:phi_scale}
\phi
=(1+\varepsilon)M_{\rm Pl}^2\,\exp\!\left[\varepsilon\ln\!\left(\frac{R}{M_{\rm Pl}^2}\right)\right],
\end{equation}
and using $\varepsilon\ll1$ together with $R\ll M_{\rm Pl}^2$ throughout the EFT regime, one has
$\varepsilon|\ln(R/M_{\rm Pl}^2)|\ll1$, hence $\phi(T_{\max})=\mathcal{O}(M_{\rm Pl}^2)$ up to $\mathcal{O}(1)$ factors.
Consequently, for $M_\ast\simeq M_{\rm Pl}$ and ${\cal O}(1)$ couplings, the spectator approximation is automatically satisfied
provided $T_{\max}\ll M_\ast$. For the benchmark values relevant here one may take $T_{\max}\sim 10^{14}\,$GeV (or more generally
$T_{\max}=T_{\rm RH}$ within observationally allowed reheating temperatures), which ensures $\delta_{\rm STB}(T)\ll1$ throughout
the baryogenesis epoch.

\subsubsection*{Baryon asymmetry}
Substituting the time–temperature relation obtained above together with $\dot\phi(t)$ in Eq.~\eqref{scalar_F(R)} into the general STB yield~\eqref{eq:asymmetry phi}, and using the Weinberg-operator decoupling condition $\Gamma_W(T_D)=H(T_D)$ to eliminate $T_D$, one finds the compact expression
\begin{equation}
\label{eq:nb_s_fR_weinberg_final}
\frac{n_b}{s}\;\simeq\;
-\frac{\hat{g}\,\mathrm{c}_{\text{STB}}\,c_s\,\varepsilon\,\phi_0\,\pi^3\,v_H^4\,(1+\varepsilon)}
{6\,M_\ast^2\,\rho_0\,\bar m_\nu^{\,2}}\,,
\end{equation}
where $\phi_0$ and $\rho_0$ are the (model-dependent) normalization constants defined in the background solution, and $\bar m_\nu^{\,2}$ denotes the flavour-invariant neutrino-mass combination controlling the $\Delta L=2$ rate induced by the Weinberg operator.

Equation~\eqref{eq:nb_s_fR_weinberg_final} makes explicit that the predicted asymmetry scales linearly with the EFT coupling $\mathrm{c}_{\text{STB}}$ and with the deviation parameter $\varepsilon$ of the $F(R)$ model. The latter is the only free parameter of the $F(R)$ model and is bounded by Big-Bang Nucleosynthesis considerations to satisfy $\varepsilon \lesssim 4\times10^{-6}$~\cite{Kusakabe:2015yaa}.
\begin{figure}[t!]
\centering
\includegraphics[width=0.49\textwidth]{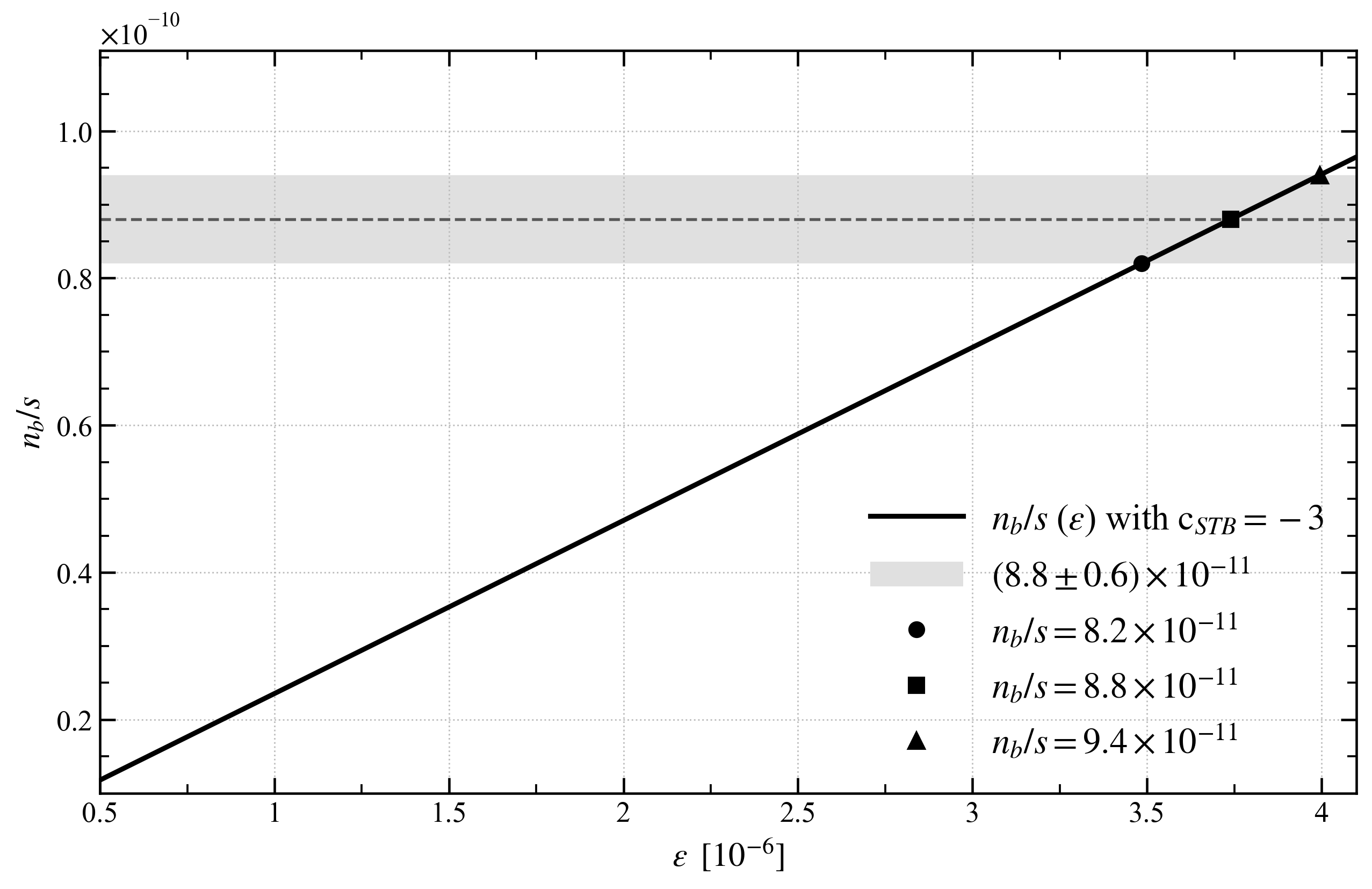}
\caption{Baryon asymmetry yield $n_b/s$ as a function of $\varepsilon$ for the model $F(R)=R^{1+\varepsilon}$, evaluated with $M_\ast=M_{\rm Pl}$, $\mathrm{c}_{\text{STB}}=-3$, and the minimal normal-ordering benchmark $\bar m_\nu^{\,2}\simeq 2.6\times10^{-3}\,\mathrm{eV}^2$. The shaded region denotes the observational interval $(8.8\pm0.6)\times10^{-11}$. The markers indicate representative intersections with the lower, central, and upper values of the band: $\varepsilon_\circ = 3.484770\times10^{-6}$ ($8.2\times10^{-11}$), $\varepsilon_\Box = 3.739751\times10^{-6}$ ($8.8\times10^{-11}$), and $\varepsilon_\triangle = ~ 3.994731\times10^{-6}$ ($9.4\times10^{-11}$).}
\label{Figure_1}
\end{figure}
The remaining particle-physics input enters through $\bar m_\nu^{\,2}\equiv \sum_{i=1}^3 m_{\nu_i}^2 = \mathrm{Tr}(m_\nu^\dagger m_\nu)$. In what follows we adopt a minimal benchmark for the light-neutrino spectrum, namely normal ordering with negligible lightest mass, in which case oscillation data fix
\begin{equation}
\bar m_\nu^{\,2}\simeq \Delta m_{21}^2+\Delta m_{31}^2 \simeq 2.6\times10^{-3}\,\mathrm{eV}^2\,,
\end{equation}
within current data~\cite{ParticleDataGroup:2024cfk,Esteban:2024eli}.
Setting $M_\ast=M_{\rm Pl}$ and $\mathrm{c}_{\text{STB}}=-3$, we display in Fig.~\ref{Figure_1} the resulting dependence of $n_b/s$ on $\varepsilon$ and indicate the intersections with the observational band.
As shown in this figure, values $\varepsilon\simeq\{3.484770,\,3.739751,\,3.994731\}\times10^{-6}$, with the decoupling temperatures $\{8.51023,8.51029,8.51034 \}\times 10^{13} \ \text{GeV}$ that satisfy the negligible backreaction consideration, reproduce the observed baryon asymmetry while remaining within the nucleosynthesis bound. Over this interval, the background expansion exponent $\eta=(1+\varepsilon)/2$ stays extremely close to its general-relativistic radiation value $\eta_{\rm GR}=1/2$, e.g.
$\eta(\varepsilon_\circ)=0.50000175$, $\eta(\varepsilon_\Box)=0.50000186$, and $\eta(\varepsilon_\triangle)=.50000199$.
This illustrates that a parametrically small deformation of GR in the gravitational sector can nevertheless provide a viable STB realization, consistent with both the measured baryon asymmetry and standard early-Universe constraints.
%%%%%%%%%%%%%%%%%%%%%%%%%%%%%%%%%%%%%%%%%%%%%%%%%%%%
\section{Summary and discussion}\label{sec:Summuary and discussion}
%%%%%%%%%%%%%%%%%%%%%%%%%%%%%%%%%%%%%%%%%%%%%%%%%%%%
In this paper, we have proposed \emph{Scalar--Tensor Baryogenesis}, a baryogenesis framework in which the effective
$C\!P$-violating bias is sourced by the gravitational scalar degrees of freedom that arise in scalar--tensor
representations of modified gravity. The mechanism is driven by derivative operators
$M_\ast^{-d}\nabla_\mu f(\phi_i)\,J^\mu_{B-L}$ which, in a homogeneous FRW background, generate an effective
chemical potential $\mu_{B-L}\propto \dot f$ and produce an equilibrium $B\!-\!L$ density while $B\!-\!L$-violating
reactions remain in equilibrium; the asymmetry freezes in at the dynamically determined decoupling temperature
$T_D$ fixed by $\Gamma_{B-L}(T_D)=H(T_D)$.

A central result of this work is structural. Whenever a modified-gravity theory admits a (local) scalar--tensor representation with a well-defined Legendre map, the gravity-sector relations $x_i=V_{\phi_i}(\phi_1,\ldots,\phi_N)$ provide an explicit \emph{on-shell/background} dictionary between generalized
``geometric'' gravitational-baryogenesis operators of the form $\nabla_\mu\mathcal{I}(x_1,\ldots,x_N)\,J^\mu_{B-L}$ and scalar--tensor couplings
$\nabla_\mu f(\phi_1,\ldots,\phi_N)\,J^\mu_{B-L}$. This shows how broad classes of generalized GB operators can be organized and interpreted within the scalar--tensor completion, while making explicit that the correspondence is local and branch-dependent and should be applied along the homogeneous background relevant for the baryogenesis
computation. STB can be viewed as a gravity-completion and organizational framework for derivative baryogenesis: the biasing scalar(s) are the gravitational degrees of freedom of the scalar--tensor representation, and geometric GB couplings admit an explicit on-shell/background translation via the Legendre relations. The correspondence comes with selection rules (branch existence, local invertibility of the Legendre map, and spectator validity) and makes $\mu_{B-L}\propto \dot f$ model-tied once $F$ is specified.

We illustrated the framework in the benchmark model $F(R)=M_{\rm Pl}^{2-2\varepsilon}R^{1+\varepsilon}$ with the dimension-five Weinberg operator as a minimal source of $B\!-\!L$ violation. In this setup the freeze-out temperature is fixed by the microscopic condition $\Gamma_W(T_D)=H(T_D)$ and we obtain viable baryogenesis for $\varepsilon=\mathcal{O}(10^{-6})$ with $T_D\simeq 8.5\times10^{13}\,$GeV, while satisfying the nucleosynthesis bound on $\varepsilon$ and keeping the expansion arbitrarily close to the GR radiation solution. Because STB operators depend explicitly on gravitational scalars, they generically source the scalar equations of motion; we quantified this effect and showed that the spectator approximation required for the background/on-shell dictionary is self-consistent throughout the baryogenesis epoch. In particular, the backreaction measure $\delta_{\rm STB}(T)$ is parametrically suppressed within the EFT regime by powers of $T/M_\ast$ and remains negligible for $M_\ast\simeq M_{\rm Pl}$ and reheating temperatures $T\in[T_D,T_{\max}]$ with $T_{\max}\ll M_\ast$ (e.g.\ $T_{\max}\sim 10^{14}\,$GeV), ensuring that the interaction biases the plasma without distorting the homogeneous solution used to compute the yield.

STB also admits a clean Einstein-frame formulation: for suitable functional choices the interaction reduces to the canonical derivative coupling $\partial_\mu\tilde\phi\,\tilde J^\mu$ familiar from spontaneous/quintessential baryogenesis, with the crucial difference that the biasing field is \emph{gravitational} in origin and its dynamics are fixed by the modified-gravity sector via the Legendre structure. This makes baryogenesis a controlled probe of the gravitational model rather than an arbitrary EFT choice. For the $R^\epsilon$ model, the Einstein-frame expressions provide a scalar--tensor representation of the same Jordan-frame theory. Because the conformal transformation reshuffles the matter couplings, the baryogenesis source terms in the two frames are not expected to coincide term-by-term; the meaningful comparison is instead at the level of the final asymmetry obtained after a consistent treatment of the transformed matter sector and freeze-out condition.

Natural directions for future work include extending the analysis to multi-invariant theories $F(x_1,\ldots,x_N)$ with genuinely multi-field couplings $f(\phi_1,\ldots,\phi_N)$, exploring alternative sources of $B\!-\!L$ violation (and their distinct freeze-out histories), and quantifying controlled departures from the spectator regime to assess backreaction corrections to the background/on-shell dictionary. It would also be interesting to embed STB in more complete early-Universe histories (inflation and reheating) and to investigate potential correlated signatures in primordial or gravitational-wave observables sourced by the same scalar--tensor
dynamics.

%%%%%%%%%%%%%%%%%%%%%%%%%%%%%%%%%%%%%%%%%%%%%%%%%%%%%%%%%%%%%%%%%%%%%%%%%%%%
\section*{Acknowledgements}
We thank Jos\'{e} Pedro Mimoso, Francisco Lobo, Miguel Pinto and specially Jess Rutschi for useful discussions that were critical for this work. This research was funded by the Fundação para a Ciência e a Tecnologia (FCT) from the research grants UIDB/04434/2020, UIDP/04434/2020.

\bibliographystyle{elsarticle-num} 
\bibliography{biblio}

@article{c75ffd80-cea5-30a0-aee9-19091a4f0a9f,
 ISSN = {00368733, 19467087},
 URL = {http://www.jstor.org/stable/24966477},
 author = {Frank Wilczek},
 journal = {Scientific American},
 number = {6},
 pages = {82--91},
 publisher = {Scientific American, a division of Nature America, Inc.},
 title = {The Cosmic Asymmetry betWeen Matter and Antimatter},
 urldate = {2024-11-20},
 volume = {243},
 year = {1980}
}

@article{Burles:2000ju,
    author = "Burles, S. and Nollett, K. M. and Turner, Michael S.",
    title = "{What is the BBN prediction for the baryon density and how reliable is it?}",
    eprint = "astro-ph/0008495",
    archivePrefix = "arXiv",
    reportNumber = "FERMILAB-PUB-00-239-A",
    doi = "10.1103/PhysRevD.63.063512",
    journal = "Phys. Rev. D",
    volume = "63",
    pages = "063512",
    year = "2001"
}

@article{WMAP:2003ivt,
    author = "Bennett, C. L. and others",
    collaboration = "WMAP",
    title = "{First year Wilkinson Microwave Anisotropy Probe (WMAP) observations: Preliminary maps and basic results}",
    eprint = "astro-ph/0302207",
    archivePrefix = "arXiv",
    doi = "10.1086/377253",
    journal = "Astrophys. J. Suppl.",
    volume = "148",
    pages = "1--27",
    year = "2003"
}

@article{Dabrowski:2008kx,
    author = "Dabrowski, Mariusz P. and Garecki, Janusz and Blaschke, David B.",
    title = "{Conformal transformations and conformal invariance in gravitation}",
    eprint = "0806.2683",
    archivePrefix = "arXiv",
    primaryClass = "gr-qc",
    doi = "10.1002/andp.200810331",
    journal = "Annalen Phys.",
    volume = "18",
    pages = "13--32",
    year = "2009"
}

@book{Birrell:1982ix,
    author = "Birrell, N. D. and Davies, P. C. W.",
    title = "{Quantum Fields in Curved Space}",
    doi = "10.1017/CBO9780511622632",
    isbn = "978-0-511-62263-2, 978-0-521-27858-4",
    publisher = "Cambridge University Press",
    address = "Cambridge, UK",
    series = "Cambridge Monographs on Mathematical Physics",
    year = "1982"
}

@article{Burles:2000zk,
    author = "Burles, Scott and Nollett, Kenneth M. and Turner, Michael S.",
    title = "{Big bang nucleosynthesis predictions for precision cosmology}",
    eprint = "astro-ph/0010171",
    archivePrefix = "arXiv",
    reportNumber = "FERMILAB-PUB-00-382-A",
    doi = "10.1086/320251",
    journal = "Astrophys. J. Lett.",
    volume = "552",
    pages = "L1--L6",
    year = "2001"
}

@article{Brans:1961sx,
    author = "Brans, C. and Dicke, R. H.",
    editor = "Hsu, Jong-Ping and Fine, D.",
    title = "{Mach's principle and a relativistic theory of gravitation}",
    doi = "10.1103/PhysRev.124.925",
    journal = "Phys. Rev.",
    volume = "124",
    pages = "925--935",
    year = "1961"
}

@article{OHanlon:1972xqa,
    author = "O'Hanlon, John",
    title = "{Intermediate-range gravity - a generally covariant model}",
    doi = "10.1103/PhysRevLett.29.137",
    journal = "Phys. Rev. Lett.",
    volume = "29",
    pages = "137--138",
    year = "1972"
}

@article{Teyssandier:1983zz,
    author = "Teyssandier, P. and Tourrenc, Ph.",
    title = "{The Cauchy problem for the R+R**2 theories of gravity without torsion}",
    doi = "10.1063/1.525659",
    journal = "J. Math. Phys.",
    volume = "24",
    pages = "2793",
    year = "1983"
}

@article{Sotiriou:2006hs,
    author = "Sotiriou, Thomas P.",
    title = "{f(R) gravity and scalar-tensor theory}",
    eprint = "gr-qc/0604028",
    archivePrefix = "arXiv",
    doi = "10.1088/0264-9381/23/17/003",
    journal = "Class. Quant. Grav.",
    volume = "23",
    pages = "5117--5128",
    year = "2006"
}

@article{Magnano:1993bd,
    author = "Magnano, Guido and Sokolowski, Leszek M.",
    title = "{On physical equivalence between nonlinear gravity theories and a general relativistic selfgravitating scalar field}",
    eprint = "gr-qc/9312008",
    archivePrefix = "arXiv",
    reportNumber = "TO-JLL-P-3-93",
    doi = "10.1103/PhysRevD.50.5039",
    journal = "Phys. Rev. D",
    volume = "50",
    pages = "5039--5059",
    year = "1994"
}

@article{Bertolami:2007gv,
    author = "Bertolami, Orfeu and Boehmer, Christian G. and Harko, Tiberiu and Lobo, Francisco S. N.",
    title = "{Extra force in f(R) modified theories of gravity}",
    eprint = "0704.1733",
    archivePrefix = "arXiv",
    primaryClass = "gr-qc",
    doi = "10.1103/PhysRevD.75.104016",
    journal = "Phys. Rev. D",
    volume = "75",
    pages = "104016",
    year = "2007"
}

@article{Goncalves:2021vci,
    author = "Gon\c{c}alves, Tiago B. and Rosa, Jo\~ao Lu\'\i{}s and Lobo, Francisco S. N.",
    title = "{Cosmology in scalar-tensor f(R,\,T) gravity}",
    eprint = "2112.02541",
    archivePrefix = "arXiv",
    primaryClass = "gr-qc",
    doi = "10.1103/PhysRevD.105.064019",
    journal = "Phys. Rev. D",
    volume = "105",
    number = "6",
    pages = "064019",
    year = "2022"
}

@article{Davoudiasl:2004gf,
    author = "Davoudiasl, Hooman and Kitano, Ryuichiro and Kribs, Graham D. and Murayama, Hitoshi and Steinhardt, Paul J.",
    title = "{Gravitational baryogenesis}",
    eprint = "hep-ph/0403019",
    archivePrefix = "arXiv",
    doi = "10.1103/PhysRevLett.93.201301",
    journal = "Phys. Rev. Lett.",
    volume = "93",
    pages = "201301",
    year = "2004"
}

@article{Li:2004hh,
    author = "Li, Hong and Li, Ming-zhe and Zhang, Xin-min",
    title = "{Gravitational leptogenesis and neutrino mass limit}",
    eprint = "hep-ph/0403281",
    archivePrefix = "arXiv",
    doi = "10.1103/PhysRevD.70.047302",
    journal = "Phys. Rev. D",
    volume = "70",
    pages = "047302",
    year = "2004"
}

@article{Lambiase:2006dq,
    author = "Lambiase, Gaetano and Scarpetta, G.",
    title = "{Baryogenesis in f(R): Theories of Gravity}",
    eprint = "astro-ph/0610367",
    archivePrefix = "arXiv",
    doi = "10.1103/PhysRevD.74.087504",
    journal = "Phys. Rev. D",
    volume = "74",
    pages = "087504",
    year = "2006"
}

@article{Lambiase:2012tn,
    author = "Lambiase, G. and Mohanty, S. and Pizza, L.",
    title = "{Consequences of f(R)-theories of gravity on gravitational leptogenesis}",
    eprint = "1212.6026",
    archivePrefix = "arXiv",
    primaryClass = "hep-ph",
    doi = "10.1007/s10714-013-1555-4",
    journal = "Gen. Rel. Grav.",
    volume = "45",
    pages = "1771--1785",
    year = "2013"
}

@article{MohseniSadjadi:2007qk,
    author = "Mohseni Sadjadi, H.",
    title = "{Multicomponent solution in modified theory of gravity}",
    eprint = "0710.3308",
    archivePrefix = "arXiv",
    primaryClass = "gr-qc",
    doi = "10.1103/PhysRevD.77.103501",
    journal = "Phys. Rev. D",
    volume = "77",
    pages = "103501",
    year = "2008"
}

@article{Bhattacharjee:2020jfk,
    author = "Bhattacharjee, Snehasish",
    title = "{Gravitational baryogenesis in extended teleparallel theories of gravity}",
    eprint = "2005.05534",
    archivePrefix = "arXiv",
    primaryClass = "gr-qc",
    doi = "10.1016/j.dark.2020.100612",
    journal = "Phys. Dark Univ.",
    volume = "30",
    pages = "100612",
    year = "2020"
}

@article{Jaybhaye:2023lgr,
    author = "Jaybhaye, Lakhan V. and Bhattacharjee, Snehasish and Sahoo, P. K.",
    title = "{Baryogenesis in f(R,Lm) gravity}",
    eprint = "2304.02482",
    archivePrefix = "arXiv",
    primaryClass = "gr-qc",
    doi = "10.1016/j.dark.2023.101223",
    journal = "Phys. Dark Univ.",
    volume = "40",
    pages = "101223",
    year = "2023"
}

@article{Baffou:2018hpe,
    author = "Baffou, E. H. and Houndjo, M. J. S. and Kanfon, D. A. and Salako, I. G.",
    title = "{$f(R,T)$ models applied to baryogenesis}",
    eprint = "1808.01917",
    archivePrefix = "arXiv",
    primaryClass = "gr-qc",
    doi = "10.1140/epjc/s10052-019-6559-0",
    journal = "Eur. Phys. J. C",
    volume = "79",
    number = "2",
    pages = "112",
    year = "2019"
}

@article{Nozari:2018ift,
    author = "Nozari, Kourosh and Rajabi, Fateme",
    title = "{Baryogenesis in $f(R,T)$ Gravity}",
    doi = "10.1088/0253-6102/70/4/451",
    journal = "Commun. Theor. Phys.",
    volume = "70",
    number = "4",
    pages = "451",
    year = "2018"
}

@article{Sahoo:2019pat,
    author = "Sahoo, P. K. and Bhattacharjee, Snehasish",
    title = "{Gravitational Baryogenesis in Non-Minimal Coupled $f(R,T)$ Gravity}",
    eprint = "1907.13460",
    archivePrefix = "arXiv",
    primaryClass = "physics.gen-ph",
    doi = "10.1007/s10773-020-04414-3",
    journal = "Int. J. Theor. Phys.",
    volume = "59",
    number = "5",
    pages = "1451--1459",
    year = "2020"
}

@article{Pereira:2024kmj,
    author = "Pereira, David S. and Lobo, Francisco S. N. and Mimoso, Jos\'e Pedro",
    title = "{Gravitational baryogenesis in energy-momentum squared gravity}",
    eprint = "2409.04623",
    archivePrefix = "arXiv",
    primaryClass = "gr-qc",
    month = "9",
    journal = "",
    year = "2024"
}

@article{Arbuzova:2023rri,
    author = "Arbuzova, Elena and Dolgov, Alexander and Dutta, Koushik and Rangarajan, Raghavan",
    title = "{Gravitational Baryogenesis: Problems and Possible Resolution}",
    eprint = "2301.08322",
    archivePrefix = "arXiv",
    primaryClass = "gr-qc",
    doi = "10.3390/sym15020404",
    journal = "Symmetry",
    volume = "15",
    number = "2",
    pages = "404",
    year = "2023"
}

@article{Cohen:1987vi,
    author = "Cohen, Andrew G. and Kaplan, David B.",
    title = "{Thermodynamic Generation of the Baryon Asymmetry}",
    reportNumber = "HUTP-87/A061",
    doi = "10.1016/0370-2693(87)91369-4",
    journal = "Phys. Lett. B",
    volume = "199",
    pages = "251--258",
    year = "1987"
}

@article{Cohen:1988kt,
    author = "Cohen, Andrew G. and Kaplan, David B.",
    title = "{SPONTANEOUS BARYOGENESIS}",
    reportNumber = "HUTP-88/A016",
    doi = "10.1016/0550-3213(88)90134-4",
    journal = "Nucl. Phys. B",
    volume = "308",
    pages = "913--928",
    year = "1988"
}

@book{Kolb:1990vq,
    author = "Kolb, Edward W.",
    title = "{The Early Universe}",
    reportNumber = "FERMILAB-BOOK-1990-01",
    doi = "10.1201/9780429492860",
    isbn = "978-0-429-49286-0, 978-0-201-62674-2",
    publisher = "Taylor and Francis",
    volume = "69",
    month = "5",
    year = "2019"
}

@article{DeFelice:2004uv,
    author = "De Felice, Antonio and Trodden, Mark",
    title = "{Baryogenesis after hyperextended inflation}",
    eprint = "hep-ph/0412020",
    archivePrefix = "arXiv",
    doi = "10.1103/PhysRevD.72.043512",
    journal = "Phys. Rev. D",
    volume = "72",
    pages = "043512",
    year = "2005"
}

@article{Sadjadi:2007dx,
    author = "Sadjadi, H. Mohseni",
    title = "{A Note on Gravitational Baryogenesis}",
    eprint = "0709.0697",
    archivePrefix = "arXiv",
    primaryClass = "gr-qc",
    doi = "10.1103/PhysRevD.76.123507",
    journal = "Phys. Rev. D",
    volume = "76",
    pages = "123507",
    year = "2007"
}

@article{DeFelice:2002ir,
    author = "De Felice, Antonio and Nasri, Salah and Trodden, Mark",
    title = "{Quintessential baryogenesis}",
    eprint = "hep-ph/0207211",
    archivePrefix = "arXiv",
    reportNumber = "SU-GP-02-7-2, SU-4252-766",
    doi = "10.1103/PhysRevD.67.043509",
    journal = "Phys. Rev. D",
    volume = "67",
    pages = "043509",
    year = "2003"
}

@article{DeFelice:2010aj,
    author = "De Felice, Antonio and Tsujikawa, Shinji",
    title = "{f(R) theories}",
    eprint = "1002.4928",
    archivePrefix = "arXiv",
    primaryClass = "gr-qc",
    doi = "10.12942/lrr-2010-3",
    journal = "Living Rev. Rel.",
    volume = "13",
    pages = "3",
    year = "2010"
}

@article{Mavromatos:2013vqa,
    author = "Mavromatos, Nick E.",
    editor = "Branco, Gustavo C. and Emmanuel-Costa, David and Gonzalez Felipe, Ricardo and Joaquim, Filipe R. and Lavoura, L. and Palomares-Ruiz, S. and Rebelo, M. Nesbitt and Romao, Jorge C. and Silva, Jo\~ao P. and Silva-Marcos, J. I.",
    title = "{Violation of CPT Invariance in the Early Universe and Leptogenesis/Baryogenesis}",
    doi = "10.1088/1742-6596/447/1/012016",
    journal = "J. Phys. Conf. Ser.",
    volume = "447",
    pages = "012016",
    year = "2013"
}

@article{McDonald:2014yfg,
    author = "McDonald, J. I. and Shore, Graham M.",
    title = "{Gravitational leptogenesis, C, CP and strong equivalence}",
    eprint = "1411.3669",
    archivePrefix = "arXiv",
    primaryClass = "hep-th",
    doi = "10.1007/JHEP02(2015)076",
    journal = "JHEP",
    volume = "02",
    pages = "076",
    year = "2015"
}

@article{Xia:2008si,
    author = "Xia, Jun-Qing and Li, Hong and Zhao, Gong-Bo and Zhang, Xinmin",
    title = "{Testing CPT Symmetry with CMB Measurements: Update after WMAP5}",
    eprint = "0803.2350",
    archivePrefix = "arXiv",
    primaryClass = "astro-ph",
    doi = "10.1086/589447",
    journal = "Astrophys. J. Lett.",
    volume = "679",
    pages = "L61--L63",
    year = "2008"
}

@article{Li:2006ss,
    author = "Li, Mingzhe and Xia, Jun-Qing and Li, Hong and Zhang, Xinmin",
    title = "{Cosmological CPT violation, baryo/leptogenesis and CMB polarization}",
    eprint = "hep-ph/0611192",
    archivePrefix = "arXiv",
    doi = "10.1016/j.physletb.2007.06.050",
    journal = "Phys. Lett. B",
    volume = "651",
    pages = "357--362",
    year = "2007"
}

@article{Mavromatos:2013boa,
    author = "Mavromatos, Nick E. and Sarkar, Sarben",
    editor = "Diosi, Lajos and Elze, Hans-Thomas and Fronzoni, Leone and Halliwell, Jonathan and Prati, Enrico and Vitiello, Giuseppe and Yearsley, James",
    title = "{CPT-violating leptogenesis induced by gravitational backgrounds}",
    doi = "10.1088/1742-6596/442/1/012020",
    journal = "J. Phys. Conf. Ser.",
    volume = "442",
    pages = "012020",
    year = "2013"
}

@article{Zhai:2020vob,
    author = "Zhai, Hua and Li, Si-Yu and Li, Mingzhe and Li, Hong and Zhang, Xinmin",
    title = "{The effects on CMB power spectra and bispectra from the polarization rotation and its correlations with temperature and E-polarization}",
    eprint = "2006.01811",
    archivePrefix = "arXiv",
    primaryClass = "astro-ph.CO",
    doi = "10.1088/1475-7516/2020/12/051",
    journal = "JCAP",
    volume = "12",
    pages = "051",
    year = "2020"
}

@article{Mavromatos:2017gyn,
    author = "Mavromatos, Nick E.",
    title = "{Models and (some) Searches for CPT Violation: From Early Universe to the Present Era}",
    doi = "10.1088/1742-6596/873/1/012006",
    journal = "J. Phys. Conf. Ser.",
    volume = "873",
    number = "1",
    pages = "012006",
    year = "2017"
}

@article{Li:2008tma,
    author = "Li, Mingzhe and Zhang, Xinmin",
    title = "{Cosmological CPT violating effect on CMB polarization}",
    eprint = "0810.0403",
    archivePrefix = "arXiv",
    primaryClass = "astro-ph",
    doi = "10.1103/PhysRevD.78.103516",
    journal = "Phys. Rev. D",
    volume = "78",
    pages = "103516",
    year = "2008"
}

@article{Barrow:2022gsu,
    author = "Barrow, J. L. and others",
    title = "{Theories and Experiments for Testable Baryogenesis Mechanisms: A Snowmass White Paper}",
    eprint = "2203.07059",
    archivePrefix = "arXiv",
    primaryClass = "hep-ph",
    month = "3",
    journal="",
    year = "2022"
}

@article{Carroll:2005dj,
    author = "Carroll, Sean M. and Shu, Jing",
    title = "{Models of baryogenesis via spontaneous Lorentz violation}",
    eprint = "hep-ph/0510081",
    archivePrefix = "arXiv",
    reportNumber = "EFI-2005-17",
    doi = "10.1103/PhysRevD.73.103515",
    journal = "Phys. Rev. D",
    volume = "73",
    pages = "103515",
    year = "2006"
}

@article{Odintsov:2016hgc,
    author = "Odintsov, S. D. and Oikonomou, V. K.",
    title = "{Gauss\textendash{}Bonnet gravitational baryogenesis}",
    eprint = "1607.00545",
    archivePrefix = "arXiv",
    primaryClass = "gr-qc",
    doi = "10.1016/j.physletb.2016.06.074",
    journal = "Phys. Lett. B",
    volume = "760",
    pages = "259--262",
    year = "2016"
}

@article{Mojahed:2024yus,
    author = "Mojahed, Martin A. and Schmitz, Kai and Xu, Xun-Jie",
    title = "{Gravitational chargegenesis}",
    eprint = "2409.10605",
    archivePrefix = "arXiv",
    primaryClass = "hep-ph",
    reportNumber = "MITP-24-074, MS-TP-24-25",
    journal = "",
    month = "9",
    year = "2024"
}

@inproceedings{Harvey:2005it,
    author = "Harvey, Jeffrey A.",
    title = "{TASI 2003 lectures on anomalies}",
    eprint = "hep-th/0509097",
    archivePrefix = "arXiv",
    reportNumber = "EFI-05-16",
    month = "9",
    year = "2005"
}

@article{Adler:1969gk,
    author = "Adler, Stephen L.",
    title = "{Axial vector vertex in spinor electrodynamics}",
    doi = "10.1103/PhysRev.177.2426",
    journal = "Phys. Rev.",
    volume = "177",
    pages = "2426--2438",
    year = "1969"
}

@article{Bell:1969ts,
    author = "Bell, J. S. and Jackiw, R.",
    title = "{A PCAC puzzle: $\pi^0 \to \gamma \gamma$ in the $\sigma$ model}",
    doi = "10.1007/BF02823296",
    journal = "Nuovo Cim. A",
    volume = "60",
    pages = "47--61",
    year = "1969"
}

@article{Bardeen:1969md,
    author = "Bardeen, William A.",
    title = "{Anomalous Ward identities in spinor field theories}",
    doi = "10.1103/PhysRev.184.1848",
    journal = "Phys. Rev.",
    volume = "184",
    pages = "1848--1857",
    year = "1969"
}

@article{Fujikawa:1979ay,
    author = "Fujikawa, Kazuo",
    title = "{Path Integral Measure for Gauge Invariant Fermion Theories}",
    reportNumber = "INS-328",
    doi = "10.1103/PhysRevLett.42.1195",
    journal = "Phys. Rev. Lett.",
    volume = "42",
    pages = "1195--1198",
    year = "1979"
}

@article{Fujikawa:1980eg,
    author = "Fujikawa, Kazuo",
    title = "{Path Integral for Gauge Theories with Fermions}",
    reportNumber = "INS-370",
    doi = "10.1103/PhysRevD.21.2848",
    journal = "Phys. Rev. D",
    volume = "21",
    pages = "2848",
    year = "1980",
    note = "[Erratum: Phys.Rev.D 22, 1499 (1980)]"
}

@article{tHooft:1976rip,
    author = "'t Hooft, Gerard",
    editor = "Shifman, Mikhail A.",
    title = "{Symmetry Breaking Through Bell-Jackiw Anomalies}",
    reportNumber = "PRINT-76-0254 (HARVARD)",
    doi = "10.1103/PhysRevLett.37.8",
    journal = "Phys. Rev. Lett.",
    volume = "37",
    pages = "8--11",
    year = "1976"
}

@article{tHooft:1976snw,
    author = "'t Hooft, Gerard",
    editor = "Shifman, Mikhail A.",
    title = "{Computation of the Quantum Effects Due to a Four-Dimensional Pseudoparticle}",
    reportNumber = "PRINT-76-0551 (HARVARD)",
    doi = "10.1103/PhysRevD.14.3432",
    journal = "Phys. Rev. D",
    volume = "14",
    pages = "3432--3450",
    year = "1976",
    note = "[Erratum: Phys.Rev.D 18, 2199 (1978)]"
}

@article{Kuzmin:1985mm,
    author = "Kuzmin, V. A. and Rubakov, V. A. and Shaposhnikov, M. E.",
    title = "{On the Anomalous Electroweak Baryon Number Nonconservation in the Early Universe}",
    reportNumber = "IC/85/8",
    doi = "10.1016/0370-2693(85)91028-7",
    journal = "Phys. Lett. B",
    volume = "155",
    pages = "36",
    year = "1985"
}

@article{Pereira:2024vdk,
    author = "Pereira, David Silva and Ferraz, Jo{\~a}o and Lobo, Francisco S. N. and Mimoso, Jos{\'e} Pedro",
    title = "{Thermodynamics of the Primordial Universe}",
    eprint = "2411.03018",
    archivePrefix = "arXiv",
    primaryClass = "gr-qc",
    doi = "10.3390/e26110947",
    journal = "Entropy",
    volume = "26",
    number = "11",
    pages = "947",
    year = "2024"
}

@article{Bodeker:2020ghk,
    author = "Bodeker, Dietrich and Buchmuller, Wilfried",
    title = "{Baryogenesis from the weak scale to the grand unification scale}",
    eprint = "2009.07294",
    archivePrefix = "arXiv",
    primaryClass = "hep-ph",
    reportNumber = "DESY 20-141, DESY-20-141",
    doi = "10.1103/RevModPhys.93.035004",
    journal = "Rev. Mod. Phys.",
    volume = "93",
    number = "3",
    pages = "035004",
    year = "2021"
}

@article{Comelli:1994di,
    author = "Comelli, D. and Pietroni, M. and Riotto, A.",
    title = "{Linear response theory approach to spontaneous baryogenesis}",
    eprint = "hep-ph/9406369",
    archivePrefix = "arXiv",
    reportNumber = "DFPD-94-TH-39, SISSA-94-82-A",
    doi = "10.1016/0927-6505(95)00021-8",
    journal = "Astropart. Phys.",
    volume = "4",
    pages = "71--86",
    year = "1995"
}

@article{Kamada:2012ht,
    author = "Kamada, Kohei and Yamaguchi, Masahide",
    title = "{Asymmetric Dark Matter from Spontaneous Cogenesis in the Supersymmetric Standard Model}",
    eprint = "1201.2636",
    archivePrefix = "arXiv",
    primaryClass = "hep-ph",
    reportNumber = "DESY-12-004",
    doi = "10.1103/PhysRevD.85.103530",
    journal = "Phys. Rev. D",
    volume = "85",
    pages = "103530",
    year = "2012"
}

@article{Harvey:1990qw,
    author = "Harvey, Jeffrey A. and Turner, Michael S.",
    title = "{Cosmological baryon and lepton number in the presence of electroweak fermion number violation}",
    reportNumber = "FERMILAB-PUB-90-049-A, EFI-90-33",
    doi = "10.1103/PhysRevD.42.3344",
    journal = "Phys. Rev. D",
    volume = "42",
    pages = "3344--3349",
    year = "1990"
}

@article{Weinberg:1979sa,
    author = "Weinberg, Steven",
    title = "{Baryon and Lepton Nonconserving Processes}",
    reportNumber = "HUTP-79-A050",
    doi = "10.1103/PhysRevLett.43.1566",
    journal = "Phys. Rev. Lett.",
    volume = "43",
    pages = "1566--1570",
    year = "1979"
}

@article{Turner:2018mwh,
    author = "Turner, Jessica and Zhou, Ye-Ling",
    title = "{Leptogenesis via Varying Weinberg Operator: the Closed-Time-Path Approach}",
    eprint = "1808.00470",
    archivePrefix = "arXiv",
    primaryClass = "hep-ph",
    reportNumber = "IPPP/18/65, FERMILAB-PUB-18-329-T",
    doi = "10.1007/JHEP01(2020)022",
    journal = "JHEP",
    volume = "01",
    pages = "022",
    year = "2020"
}

@article{Carloni:2005Rn,
  author        = {Carloni, S. and Dunsby, P. K. S. and Capozziello, S. and Troisi, A.},
  title         = {Cosmological dynamics of $R^n$ gravity},
  journal       = {Classical and Quantum Gravity},
  volume        = {22},
  number        = {22},
  pages         = {4839--4868},
  year          = {2005},
  doi           = {10.1088/0264-9381/22/22/011},
  eprint        = {gr-qc/0410046},
  archivePrefix = {arXiv},
  primaryClass  = {gr-qc}
}

@article{CapozzielloDeFelice:2008Noether,
  author        = {Capozziello, Salvatore and De Felice, Antonio},
  title         = {$f(R)$ cosmology from Noether's symmetry},
  journal       = {Journal of Cosmology and Astroparticle Physics},
  volume        = {2008},
  number        = {08},
  pages         = {016},
  year          = {2008},
  doi           = {10.1088/1475-7516/2008/08/016},
  eprint        = {0804.2163},
  archivePrefix = {arXiv},
  primaryClass  = {gr-qc}
}

@article{ParkHwangNoh:2011Scaling,
  author        = {Park, Chan-Gyung and Hwang, Jai-chan and Noh, Hyerim},
  title         = {Constraints on a $f(R)$ gravity dark energy model with early scaling evolution},
  journal       = {Journal of Cosmology and Astroparticle Physics},
  volume        = {2011},
  number        = {09},
  pages         = {038},
  year          = {2011},
  doi           = {10.1088/1475-7516/2011/09/038},
  eprint        = {1012.1662},
  archivePrefix = {arXiv},
  primaryClass  = {astro-ph.CO}
}

@article{Zakharov:2006SSRn,
  author        = {Zakharov, A. F. and Nucita, A. A. and De Paolis, F. and Ingrosso, G.},
  title         = {Solar system constraints on $R^n$ gravity},
  journal       = {Physical Review D},
  volume        = {74},
  pages         = {107101},
  year          = {2006},
  doi           = {10.1103/PhysRevD.74.107101},
  eprint        = {astro-ph/0611051},
  archivePrefix = {arXiv},
  primaryClass  = {astro-ph}
}

@article{CapozzielloCardoneTroisi:2007LSB,
  author        = {Capozziello, S. and Cardone, V. F. and Troisi, A.},
  title         = {Low surface brightness galaxy rotation curves in the low energy limit of $R^n$ gravity: no need for dark matter?},
  journal       = {Monthly Notices of the Royal Astronomical Society},
  volume        = {375},
  number        = {4},
  pages         = {1423--1440},
  year          = {2007},
  doi           = {10.1111/j.1365-2966.2007.11401.x},
  eprint        = {astro-ph/0603522},
  archivePrefix = {arXiv},
  primaryClass  = {astro-ph}
}

@article{FrigerioMartinsSalucci:2007RC,
  author        = {Frigerio Martins, Christiane and Salucci, Paolo},
  title         = {Analysis of rotation curves in the framework of $R^n$ gravity},
  journal       = {Monthly Notices of the Royal Astronomical Society},
  volume        = {381},
  number        = {3},
  pages         = {1103--1108},
  year          = {2007},
  doi           = {10.1111/j.1365-2966.2007.12273.x},
  eprint        = {astro-ph/0703243},
  archivePrefix = {arXiv},
  primaryClass  = {astro-ph}
}

@article{Capozziello:2015yza,
    author = "Capozziello, Salvatore and De Laurentis, Mariafelicia and Farinelli, Ruben and Odintsov, Sergei D.",
    title = "{Mass-radius relation for neutron stars in f(R) gravity}",
    eprint = "1509.04163",
    archivePrefix = "arXiv",
    primaryClass = "gr-qc",
    doi = "10.1103/PhysRevD.93.023501",
    journal = "Phys. Rev. D",
    volume = "93",
    number = "2",
    pages = "023501",
    year = "2016"
}

@article{SotiriouFaraoni:2010RMP,
  author        = {Sotiriou, Thomas P. and Faraoni, Valerio},
  title         = {$f(R)$ theories of gravity},
  journal       = {Reviews of Modern Physics},
  volume        = {82},
  pages         = {451--497},
  year          = {2010},
  doi           = {10.1103/RevModPhys.82.451},
  eprint        = {0805.1726},
  archivePrefix = {arXiv},
  primaryClass  = {gr-qc}
}

@article{Kusakabe:2015yaa,
    author = "Kusakabe, Motohiko and Koh, Seoktae and Kim, K. S. and Cheoun, Myung-Ki",
    title = "{Corrected constraints on big bang nucleosynthesis in a modified gravity model of $f(R) \propto R^n$}",
    eprint = "1506.08859",
    archivePrefix = "arXiv",
    primaryClass = "astro-ph.CO",
    doi = "10.1103/PhysRevD.91.104023",
    journal = "Phys. Rev. D",
    volume = "91",
    pages = "104023",
    year = "2015"
}

@article{Pascoli:2018cqk,
    author = "Pascoli, Silvia and Turner, Jessica and Zhou, Ye-Ling",
    title = "{Leptogenesis via a varying Weinberg operator: a semi-classical approach}",
    eprint = "1808.00475",
    archivePrefix = "arXiv",
    primaryClass = "hep-ph",
    reportNumber = "IPPP/18/66, FERMILAB-PUB-18-330-T",
    doi = "10.1088/1674-1137/43/3/033101",
    journal = "Chin. Phys. C",
    volume = "43",
    number = "3",
    pages = "033101",
    year = "2019"
}

@article{ParticleDataGroup:2024cfk,
    author = "Navas, S. and others",
    collaboration = "Particle Data Group",
    title = "{Review of particle physics}",
    doi = "10.1103/PhysRevD.110.030001",
    journal = "Phys. Rev. D",
    volume = "110",
    number = "3",
    pages = "030001",
    year = "2024"
}

@article{Esteban:2024eli,
    author = "Esteban, Ivan and Gonzalez-Garcia, M. C. and Maltoni, Michele and Martinez-Soler, Ivan and Pinheiro, Jo{\~a}o Paulo and Schwetz, Thomas",
    title = "{NuFit-6.0: updated global analysis of three-flavor neutrino oscillations}",
    eprint = "2410.05380",
    archivePrefix = "arXiv",
    primaryClass = "hep-ph",
    reportNumber = "IFT-UAM/CSIC-24-140, YITP-SB-2024-24, IPPP/24/64, IPPP/24/64, IFT-UAM/CSIC-24-140, YITP-SB-2024-24",
    doi = "10.1007/JHEP12(2024)216",
    journal = "JHEP",
    volume = "12",
    pages = "216",
    year = "2024"
}

@article{Pereira:2025flo,
    author = "Pereira, David S. and Lobo, Francisco S. N. and Mimoso, Jos{\'e} P.",
    title = "{Baryon asymmetry from higher-order matter contributions in gravity}",
    eprint = "2504.21504",
    archivePrefix = "arXiv",
    primaryClass = "gr-qc",
    doi = "10.1016/j.physletb.2025.139521",
    journal = "Phys. Lett. B",
    volume = "866",
    pages = "139521",
    year = "2025"
}

@article{Cruz:2025fuk,
    author = "Cruz, Daniel F. P. and Pereira, David S. and Lobo, Francisco S. N.",
    title = "{Gravitational baryogenesis in $f(T,L_m)$ gravity}",
    eprint = "2509.17218",
    archivePrefix = "arXiv",
    primaryClass = "gr-qc",
    month = "9",
    year = "2025"
}

@article{Barrow:1988xh,
    author = "Barrow, John D. and Cotsakis, S.",
    title = "{Inflation and the Conformal Structure of Higher Order Gravity Theories}",
    doi = "10.1016/0370-2693(88)90110-4",
    journal = "Phys. Lett. B",
    volume = "214",
    pages = "515--518",
    year = "1988"
}

@book{Will:2018bme,
    author = "Will, Clifford M.",
    title = "{Theory and Experiment in Gravitational Physics}",
    isbn = "978-1-108-67982-4, 978-1-107-11744-0",
    publisher = "Cambridge University Press",
    month = "9",
    year = "2018"
}

@article{Odintsov:2016apy,
    author = "Odintsov, S. D. and Oikonomou, V. K.",
    title = "{Loop Quantum Cosmology Gravitational Baryogenesis}",
    eprint = "1610.02533",
    archivePrefix = "arXiv",
    primaryClass = "gr-qc",
    doi = "10.1209/0295-5075/116/49001",
    journal = "EPL",
    volume = "116",
    number = "4",
    pages = "49001",
    year = "2016"
}

@article{Oikonomou:2016jjh,
    author = "Oikonomou, V. K. and Saridakis, Emmanuel N.",
    title = "{$f(T)$ gravitational baryogenesis}",
    eprint = "1607.08561",
    archivePrefix = "arXiv",
    primaryClass = "gr-qc",
    doi = "10.1103/PhysRevD.94.124005",
    journal = "Phys. Rev. D",
    volume = "94",
    number = "12",
    pages = "124005",
    year = "2016"
}

\end{document}